\documentclass{emulateapj}
\usepackage{apjfonts}
\usepackage{graphicx}
\usepackage{amsmath}
     \usepackage{dpfloat}

\usepackage{longtable}

\usepackage[toc,page]{appendix}

\usepackage{color}

\def\kms{\ifmmode{\rm km\thinspace s^{-1}}\else km\thinspace s$^{-1}$\fi}

\shortauthors{Rappaport et al.~2016}
\shorttitle{Quintuple Star System}

\begin{document}

%
\def\ltsima{$\; \buildrel < \over \sim \;$}
\def\lsim{\lower.5ex\hbox{\ltsima}}
\def\gtsima{$\; \buildrel > \over \sim \;$}
\def\gsim{\lower.5ex\hbox{\gtsima}}
\def\teff{$T_\mathrm{eff}$}                 
 \def\vsini{\hbox{$v$\,sin\,$i_{\star}$}}    
  \def\cms2{\hbox{\,cm\,s$^{-2}$}}

%

\bibliographystyle{apj}

\title{A Quintuple Star System Containing Two Eclipsing Binaries}

\author{
S.~Rappaport\altaffilmark{1}, 
H.~Lehmann\altaffilmark{2},
B.~Kalomeni\altaffilmark{1,3},
T.~Borkovits\altaffilmark{4},
D.~Latham\altaffilmark{5},
A.~Bieryla\altaffilmark{5},
H.~Ngo\altaffilmark{6},
D.~Mawet\altaffilmark{7},
S.~Howell\altaffilmark{8},
E.~ Horch\altaffilmark{9}
T. L.~Jacobs\altaffilmark{10},
D.~LaCourse\altaffilmark{11},
\'{A}.~ S\'{o}dor\altaffilmark{12}, 
A.~Vanderburg\altaffilmark{5},
K.~Pavlovski\altaffilmark{13}
}

\altaffiltext{1}{Department of Physics, and Kavli Institute for
  Astrophysics and Space Research, Massachusetts Institute of
  Technology, Cambridge, MA 02139, USA, sar@mit.edu}
  
\altaffiltext{2}{Th\"{u}ringer Landessternwarte Tautenburg, Sternwarte 5, D-07778 Tautenburg, Germany; lehm@tls-tautenburg.de}

\altaffiltext{3}{Department of Astronomy and Space Sciences, Ege University, 35100, \.Izmir, Turkey; belinda.kalomeni@ege.edu.tr}

\altaffiltext{4}{Baja Astronomical Observatory of Szeged University, H-6500 Baja, Szegedi \'{u}t, Kt. 766, Hungary; borko@electra.bajaobs.hu}

\altaffiltext{5}{Harvard-Smithsonian Center for Astrophysics, 60 Garden Street, Cambridge, MA 02138 USA}

\altaffiltext{6}{California Institute of Technology, Division of Geological and Planetary Sciences, 1200 E California Blvd MC 150-21, Pasadena,
CA 91125, USA; hngo@caltech.edu}

\altaffiltext{7}{California Institute of Technology, Astronomy Dept. MC 249-17, 1200 E. California Blvd., Pasadena, CA 91125, USA; dmawet@astro.caltech.edu}

\altaffiltext{8}{Kepler \& K2 Missions, NASA Ames Research Center, PO Box 1, M/S 244-30, Moffett Field, CA 94035}

\altaffiltext{9}{Department of Physics, Southern Connecticut State University, New Haven, CT, 06515}

\altaffiltext{10}{12812 SE 69th Place Bellevue, WA 98006; tomjacobs128@gmail.com}

\altaffiltext{11}{7507 52nd Place NE Marysville, WA 98270; daryll.lacourse@gmail.com}

\altaffiltext{12}{Konkoly Observatory, MTA CSFK, Konkoly Thege M. \'ut 15--17, 1121 Budapest, Hungary; sodor@konkoly.hu}

\altaffiltext{13}{Department of Physics, University of Zagreb, Bijenicka cesta 32, 10000 Zagreb, Croatia; pavlovski@phy.hr}

\slugcomment{Submitted to {\it Monthly Notices of the Royal Astronomical Society}, 2016 May 20; Revised 2016, June 16}

\begin{abstract}

We present a quintuple star system that contains two eclipsing binaries.  The unusual architecture includes two stellar images separated by $11''$ on the sky: EPIC 212651213 and EPIC 212651234.  The more easterly image (212651213) actually hosts both eclipsing binaries which are resolved within that image at $0.09''$, while the westerly image (212651234) appears to be single in adaptive optics (AO), speckle imaging, and radial velocity (RV) studies.  The `A' binary is circular with a 5.1-day period, while the `B' binary is eccentric with a 13.1-day period.  The $\gamma$ velocities of the A and B binaries are different by $\sim$10 km s$^{-1}$.  That, coupled with their resolved projected separation of $0.09''$, indicates that the orbital period and separation of the `C' binary (consisting of A orbiting B) are $\simeq 65$ years and $\simeq 25$ AU, respectively, under the simplifying assumption of a circular orbit. Motion within the C orbit should be discernible via future RV, AO, and speckle imaging studies within a couple of years.  The C system (i.e., 212651213) has a radial velocity and proper motion that differ from that of 212651234 by only $\sim$1.4 km s$^{-1}$ and $\sim$3 mas yr$^{-1}$.  This set of similar space velocities in 3 dimensions strongly implies that these two objects are also physically bound, making this at least a quintuple star system.

\end{abstract}


\section{Introduction}

The {\em Kepler} mission, with its exquisite photometric precision (Borucki et al.~2010; Batalha et al.~2011) has revolutionized the study of stars in general, and binary stars, in particular.  There were some 3000 binaries discovered in the {\em Kepler} main field (Slawson et al.~2011; Matijevi\v{c} et al.~2012; Kirk et al.~2016), and there is a growing collection of binaries that have been found to date in the 2-wheel extension of the {\em Kepler} mission, called `K2' (Howell et al.~2014; LaCourse et al.~2015). Among this impressive collection of mostly eclipsing binaries, some 220 triple stars have been discovered (Rappaport et al.~2013; Conroy et al.~2014; Borkovits et al.~2015; Borkovits et al.~2016), mostly through eclipse timing variations, but some via 3rd-body eclipses.  In addition to the large sample of triple-star systems, a number of higher-order multiple star systems have also been discovered using {\em Kepler} data plus follow-up ground-based observations (quadruple 4247791: Lehmann et al.~2012; quadruple 7177553: Lehmann et al.~2016; and quintuple KIC 4150611/HD 181469: Shibahashi \& Kurtz 2012, and references therein; Prsa et al.~2016).  Other quintuple systems, not found with {\em Kepler}, include: 1SWASP J093010.78+533859.5 (Lohr et al.~2015); the young B-star quintuple HD 27638 (Torres 2006); HD 155448 (Sch\"utz et al.~2011); 14 Aurigae (Barstow et al.~2001); $\sigma^2$ Coronae Borealis (Raghavan et al.~2009); GG Tau (Di Folco et al.~2014), HIP 28790/28764 and HIP 64478 (Tokovinin 2016), and V994 Her (Zasche \& Uhla\v{r} 2013).  Aside from being intrinsically fascinating systems, we can learn the most about the formation and evolution of multiple star systems if the constituent stars are close enough to interact dynamically on timescales of less than a few years.

In this work we report the first quintuple star system found in the K2 fields, and one of a relatively few that contain two eclipsing and spectroscopic binaries.  This work is organized as follows.  In Sect.~\ref{sec:K2} we present the K2 Field 6 data in which the two eclipsing binaries, `A' and `B', were discovered to be in the same stellar image.  Our optical radial velocity studies are discussed in Sect.~\ref{sec:RV}, and Keplerian orbits are fit to the RV data in Sect.~\ref{sec:RVorbits}.  We evaluate the full set of parameters for the `A' and `B' binaries in Sect.~\ref{sec:MC_eval} using a Monte Carlo evaluation process.  The spectra are analyzed in Sect.~\ref{sec:decomp} to further confirm our parameter determinations for binaries `A' and `B'.  We consider the tidal status of binaries `A' and `B' in Sect.~\ref{sec:tides}.   In an effort to further probe the structure of the system we obtained adaptive-optics and speckle-interferometric images of the system; these are presented in Sect.~\ref{sec:AO}.  The photometric distance to the target is derived in Sect.~\ref{sec:dist}.  We discuss in Sect.~\ref{sec:binaryC} the constraints that can be set on `binary C' which is composed of binaries `A' and `B' that are gravitationally bound to each other.  Binary `D'  comprised of binary `C' in orbit about the apparently single star EPIC 212651234, is discussed in Sect.~\ref{sec:Dbinary}.  We investigate the single star EPIC 212651234 in Sect.~\ref{sec:E1234}.  Finally, we summarize our results in Sect.~\ref{sec:concl}.

\section{K2 Observations}
\label{sec:K2}

As part of our ongoing search for eclipsing binaries, we downloaded all available K2 Extracted Lightcurves common to Campaign Field 6 from the MAST\footnote{\url{http://archive.stsci.edu/k2/data\_search/search.php}}. The flux data from all 28,000 targets were high-pass filtered with a cut-on frequency of 1
day. The data files were then Fourier transformed to facilitate a search for objects with periodic signatures. The folded lightcurves of targets with significant peaks in their FFTs were then examined by eye to look for unusual objects with periodic features.  Unfolded and normalized light curves for targets lacking significant peaks in their FFTs were also examined by eye to look for aperiodic or quasi-periodic signatures. 

\begin{figure}[t]
\begin{center}
\includegraphics[width=0.48 \textwidth]{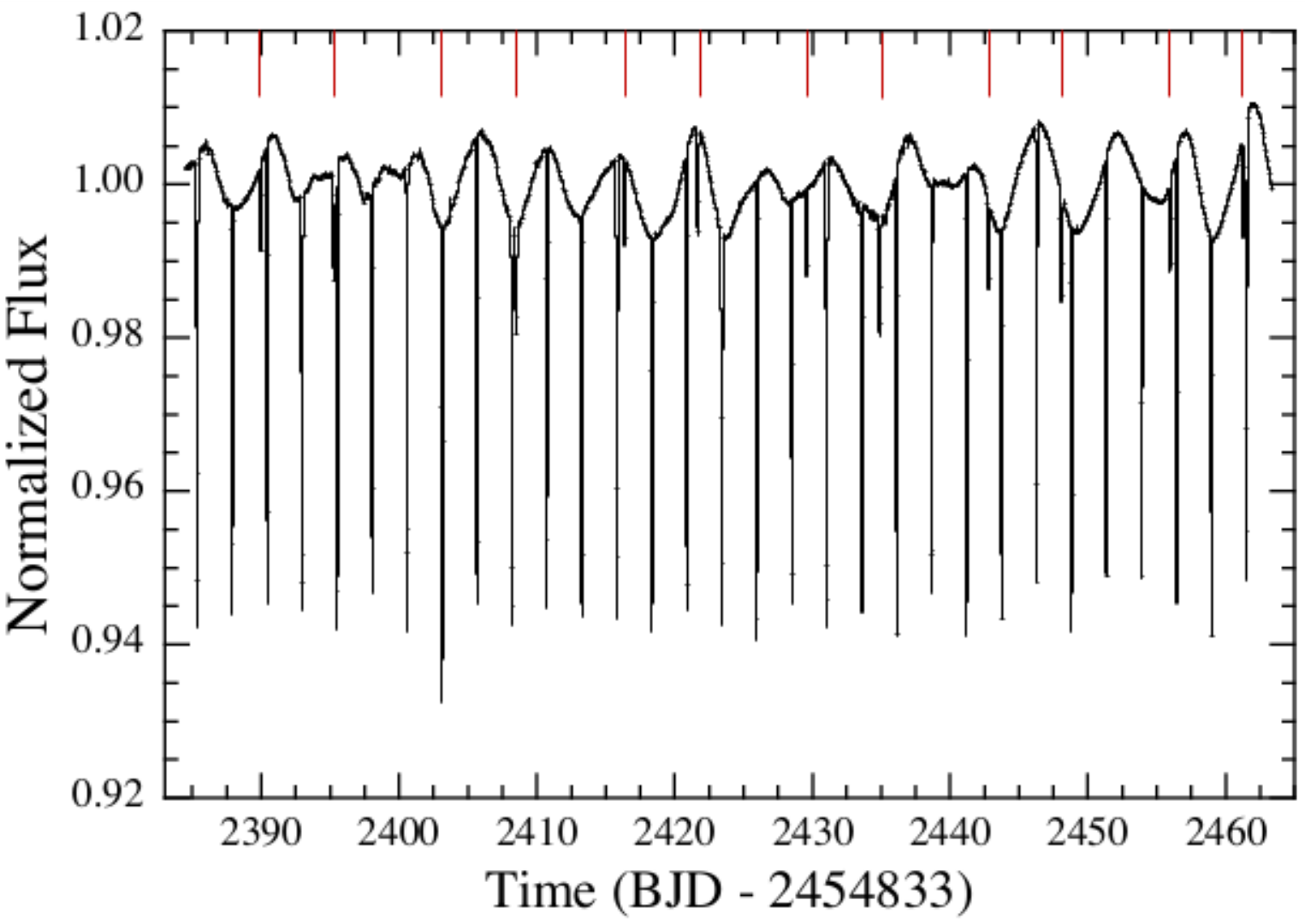}
\caption{K2 flux data for E1213.  The two eclipses of the 5-day `A' binary are readily visible.  The short red vertical lines mark the two unequally spaced eclipses of the 13-day eccentric `B' binary.}
\label{fig:rawLC}
\end{center}
\end{figure}

One object that stood out was EPIC 212651213 (hereafter `E1213') which exhibited eclipses with a 5-day period, but where the folded light curve also showed the presence of eclipses that did not fit this period.  The raw light curve is shown in Fig.~\ref{fig:rawLC} and the two nearly equal eclipses of the 5-day orbital period are readily apparent.  However, there is also another set of unequally spaced, more shallow eclipses whose locations are marked in the figure with short vertical red lines to guide the eye.  These correspond to a different eclipsing eccentric binary with a 13-day period.  

\begin{figure}[h]
\begin{center}
\includegraphics[width=0.48 \textwidth]{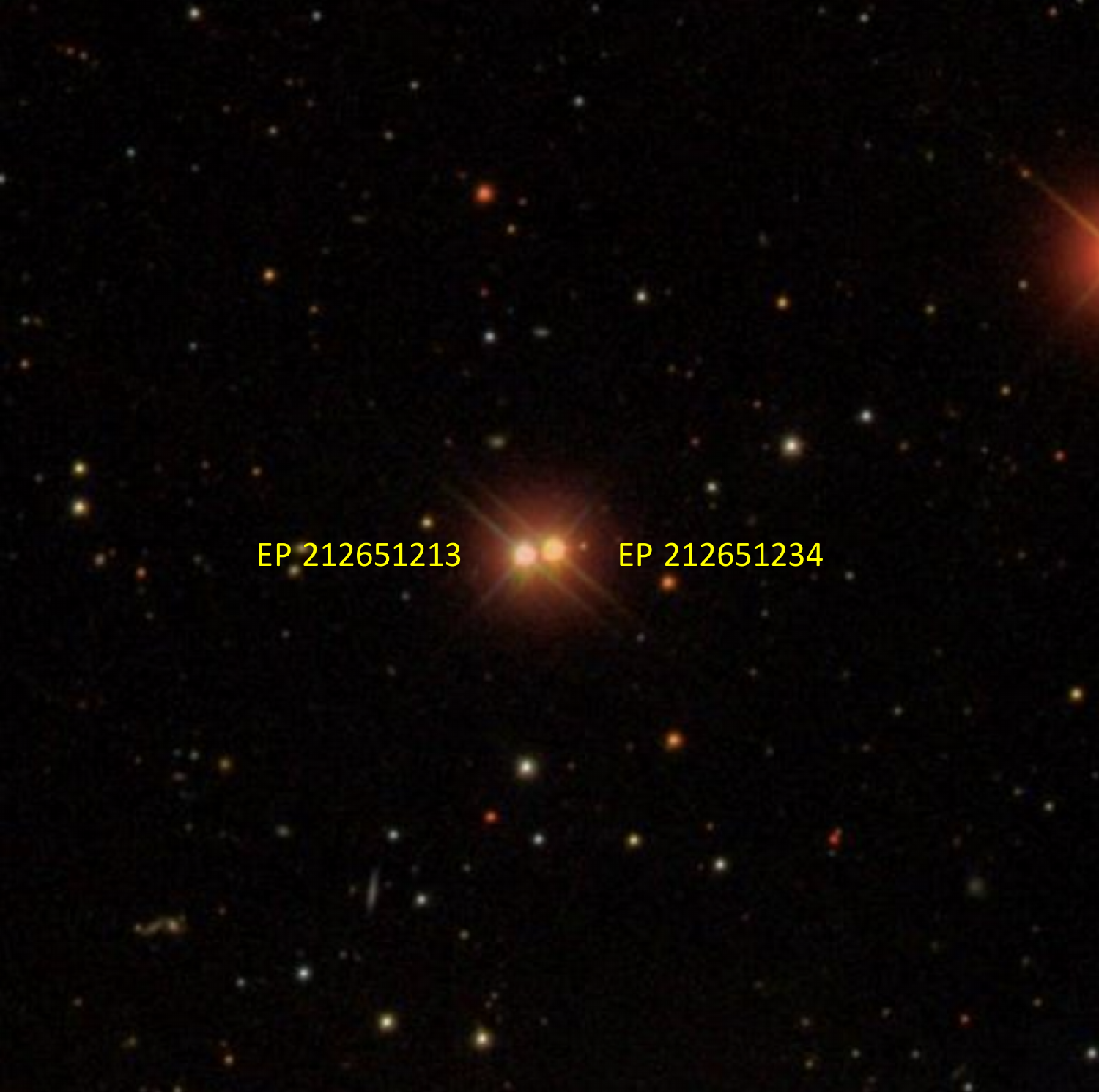}
\caption{SDSS image showing E1213 and E1234.  The image is $7' \times 7'$, and shows the two images, of roughly the same magnitude, as an obviously physically related pair.  In fact, the two objects have radial velocities and proper motions that are the same to within $\sim$1.7 km s$^{-1}$ and 3 mas yr$^{-1}$, respectively.  North is up and East is to the left.}
\label{fig:SDSS}
\end{center}
\end{figure}

We also noticed that the target EPIC 212651234 (hereafter `E1234') exhibited the same two sets of eclipses, but with reduced amplitudes.  The SDSS image of the sky in the vicinity of these two targets (see Fig.~\ref{fig:SDSS}) shows two comparably bright stellar images that are separated by $\sim$11$''$. The properties of these two stars are summarized in Table \ref{tbl:mags}.  Investigation of the Target Pixel Files for E1213 and E1234 with Guest Observer software PyKE\footnote{\url{http://keplerscience.arc.nasa.gov/software.html\#pyke}} (Still \& Barclay 2012)\footnote{\url{http://adsabs.harvard.edu/abs/2012ascl.soft08004S}}
confirmed that the location of the photometric aperture for each of these two targets slightly overlaps the other star.  By narrowing these apertures and performing a series of independent extractions on each target we were able to tentatively conclude that the source of {\em both} eclipsing binary signals is, in fact, E1213. (We have subsequently confirmed this with narrow-slit and fiber-fed spectroscopic observations -- see Sect.~\ref{sec:RV}.)

We present the folded lightcurves for each of the eclipsing binaries in Fig.~\ref{fig:folds}.  We used the Ames K2 PDC data, which have been corrected for the dilution by the other star.  In each case, we subtracted the folded light curve of the other eclipsing binary in such a way as to effectively `clean' the data of the modulations due to the other binary.  To accomplish this, the folded data used for the subtraction was produced while eliminating the phases around the other binary's eclipses.

The eclipse timing analysis, for determining orbital periods, phase zero of the eclipses, and relative eclipse separations, was carried out as described in Borkovits et al.~(2015) and Borkovits et al.~(2016). This includes allowance for local trends in the out-of-eclipse regions near the eclipses.

The 5.077-day binary (top panel of Fig.~\ref{fig:folds}), hereafter designated as `Binary A', has two nearly equal depth eclipses that are separated by very close to half an orbital cycle.  The fitted fractional separation between eclipses differs by only $325 \pm187$ ppm from half an orbital cycle, and we can then utilize the approximate expression (good to 2nd order in eccentricity $e$):
\begin{equation}
e\,\cos \omega_A \simeq \frac{\pi}{2} \left[\frac{1}{2} - \frac{\Delta t_{(II,I)}}{P_{\rm orb}} \right]
\label{eqn:ecom}
\end{equation}
to say that $e \cos \omega_A \lesssim 0.0008$, which is indicative of a rather circular orbit unless the longitude of periastron, $\omega$, for the binary is within $\sim$$1/2^\circ$ of $90^\circ$ or $270^\circ$.  In this expression $\Delta t_{(II,I)}/P_{\rm orb}$ is the fractional time difference between the two closely spaced eclipses. 
We can also utilize information from the relative widths of the two eclipses, $w_1$ and $w_2$, to find a measure of $e\,\sin \omega$.  For small $e$ and arbitrary $\omega$:
\begin{equation}
e\,\sin \omega \simeq \frac{(1-w_2/w_1)}{(1+w_2/w_1)}
\label{eqn:esom}
\end{equation}
From the K2 photometry, we determine that $w_2/w_1 = 0.996 \pm 0.014$.  Therefore, $e_A \, \sin \omega_A = 0.002 \pm 0.02$.  Thus, based on the limits obtained from Eqns.~(\ref{eqn:ecom}) and (\ref{eqn:esom}) we can constrain the orbital eccentricity of the ``A'' binary to be
$$e_A \lesssim 0.02$$

Finally, we note the out-of-eclipse modulation with a $\sim$1\% amplitude that is in phase with the binary orbit, but is not due to ellipsoidal light variations (which would have minima at the eclipses), and we ascribe this modulation to star-spots on one or both stars of the ``A'' binary.  Figure~\ref{fig:folds} also shows that the apparently irregular variations seen in the out-of-eclipse regions in Fig.~\ref{fig:rawLC} could be well modeled in this way.

In the bottom panel of Fig.~\ref{fig:folds} the 13.1947-day eccentric binary is more clearly revealed.  For this set of eclipses we find the closely spaced pair to be separated by 0.4019 ($\pm 0.0002$) orbital cycles.  From Eqn.~\ref{eqn:ecom} we find $e_B \, \cos \omega_B = 0.154$.  In Sect.~\ref{sec:RVorbits} we are able to compare this to the results we find by directly measuring $e_B$ and $\omega_B$ from RV studies, and find excellent agreement.

\begin{table}
\centering
\caption{Properties of the E1213/1234 System}
\begin{tabular}{lcc}
\hline
\hline
Parameter &
212651213 &
212651234 \\
\hline
RA (J2000) & 13:55:43.464  & 13:55:42.762 \\  
Dec (J2000) &  $-0$9:25:05.92 & $-0$9:25:04.02 \\  
$K_p$ & 10.80 & 11.14 \\
$u^a$ &  14.21  & 14.63 \\
$B^b$ & 11.77 & 12.32 \\
$V^b$ & 10.98 & 11.31 \\
$r'^c$ & 11.08 & 11.24 \\
J$^d$ & 9.93 & 9.74 \\
H$^d$ & 9.59 & 9.27 \\
K$^d$ & 9.54 & 9.18 \\
W1$^e$ & 9.36 & 9.10 \\
W2$^e$ & 9.40 & 9.21 \\
W3$^e$ & 9.30 & 9.09 \\
W4$^e$ & 8.59 & 8.69 \\
Distance (pc)$^f$ & $260 \pm 50$ &  $260 \pm 50$ \\
$V_{\rm rad}$ (km/sec) & $-13.5 \pm 0.5$$^g$ & $-15.0 \pm 0.5$ \\
$\mu_\alpha$ (mas ~${\rm yr}^{-1}$)$^h$ & $-29.7 \pm 1.3$ & $-32.6 \pm 1.9$ \\
$\mu_\delta$ (mas ~${\rm yr}^{-1}$)$^h$ & $-10.9 \pm 1.9$ & $-12.6 \pm 1.5$ \\
\hline
\label{tbl:mags}
\end{tabular}

{\bf Notes.} (a) Taken from the SDSS image (Ahn et al.~2012). (b) From APASS (\url{https://www.aavso.org/apass}). (c) Carlsberg Meridian Catalog (\url{http://svo2.cab.inta-csic.es/vocats/cmc15/}).  (d) 2MASS catalog (Skrutskie et al.~2006).  (e) WISE point source catalog (Cutri et al.~2013). (f) Based on photometric parallax only.  (g) The mass-weighted mean of the $\gamma$ velocities for the `A' and `B' binaries (see text). (h) From UCAC4 (Zacharias et al.~2013), and Huber et al.~(2015).
\end{table}

\vspace{0.5cm}

\section{Radial Velocity Observations}
\label{sec:RV}

We have carried out 27 radial velocity measurements of E1213 on 25 different nights.  Eight of these were carried out with the TRES spectrograph at the Fred Lawrence Whipple Observatory, 16 were obtained at Th\"uringer Landessternwarte Tautenburg (TLS), and 3 were acquired at the Konkoly Observatory of the Hungarian Academy of Sciences.  Here we briefly describe the spectroscopic observations at each facility. In addition, we have obtained 11 spectra of E1234.

\subsection{RV observations with the TLS Spectrograph}
\label{sec:TLS}
  
Spectra of E1213 and E1234 were taken using the Coud\'e-Echelle
Spectrograph on the 2-m Alfred Jensch Telescope at TLS.  The spectrograph was used with a 
projected slit width of 2 arcsec providing a resolving power of R = 30,000.
We obtained 16 spectra of E1213 and three spectra of E1234 with
40 minutes of exposure time each.

Spectrum reduction was done using standard ESO-MIDAS packages and included
bias and stray-light subtraction, filtering of cosmic-ray events, flat fielding 
using a halogen lamp, optimum order subtraction, wavelength calibration using 
a ThAr lamp, normalization to the local continuum, and merging of spectral
orders. Small shifts in the instrumental zero point were removed from each 
spectrum using a large number of telluric O$_2$ lines as reference. The 
long-time RV accuracy reached with this spectrograph and reduction procedure 
depends on spectral type and $v\sin{i}$ of the observed star and is about 
100 ms$^{-1}$ for sharp-lined, solar-type SB1 stars. 

\subsection{RV observations with the TRES spectrograph}
\label{sec:TRESS}
  
Spectra of E1213 and E1234 were taken using the Tillinghast Reflector Echelle Spectrograph (TRES; Szentgyorgyi \& Fur\'esz 2007), on the 1.5 m telescope at the Fred Lawrence Whipple Observatory (FLWO) on Mt.~Hopkins, Arizona. TRES is a fiber-fed optical spectrograph with a resolving power of R = 44,000. We obtained 8 spectra of E1213 between UT 2016-02-21 and UT 2016-03-23 with a typical exposure time between 660-1200 seconds, and an average signal-to-noise ratio (S/N) per resolution element of 38. E1234 was observed 6 times between UT 2016-02-21 and UT 2016-05-03 with the exposure time ranging between 360-1500 seconds, and an average S/N of 28. The spectra were extracted following the procedures reported by Buchhave et al.~(2010). 

\subsection{RV observations at the Konkoly Observatory}
\label{sec:KO}

E1213 and E1234 were observed during three nights in March 2016 with the
fiber-fed ACE spectrograph attached to the 1-m RCC telescope of the
Konkoly Observatory at Piszk\'es-tet\H o, Hungary. The instrument
covers the 0.415 $\mu$m \,--\,0.915 $\mu$m\, wavelength range with a resolution of
R\,=\,20,000. The fiber entrance has a sky-projected diameter of
$\approx\,2.5''$, which assured complete separation of the light from
the two stellar images of E1213 and E1234. The light of a
Thorium--Argon (ThAr) lamp can be projected on the entrance of the
same fiber that is used for stellar observations for precise
wavelength calibration.

The individual 1800-s exposures had S/N\,$\approx$\,25 per pixel around
0.55 $\mu$m. The RV measurements were averaged for 3\,--\,4 consecutive
exposures, with the exception of the night 2\,457\,466, when only one
exposure of E1213 was obtained.

The spectra were reduced using IRAF\footnote{IRAF is distributed by
the National Optical Astronomy Observatories, which are operated by
the Association of Universities for Research in Astronomy, Inc., under
cooperative agreement with the National Science Foundation.} standard
tasks including bias, dark, and flat-field corrections, aperture
extraction, wavelength calibration (using ThAr exposures) and
barycentric correction. The normalisation, cosmic-ray filtering, order
merging and cross-correlation were performed by in-house developed
programs of \'AS.

The systematic errors introduced by the data processing and the
stability of the wavelength calibration system was found to be better
than 0.36\,km\,s$^{-1}$, based on RV standard observations (Derekas et
al.~2016).

\begin{figure}[h]
\begin{center}
\includegraphics[width=0.476 \textwidth]{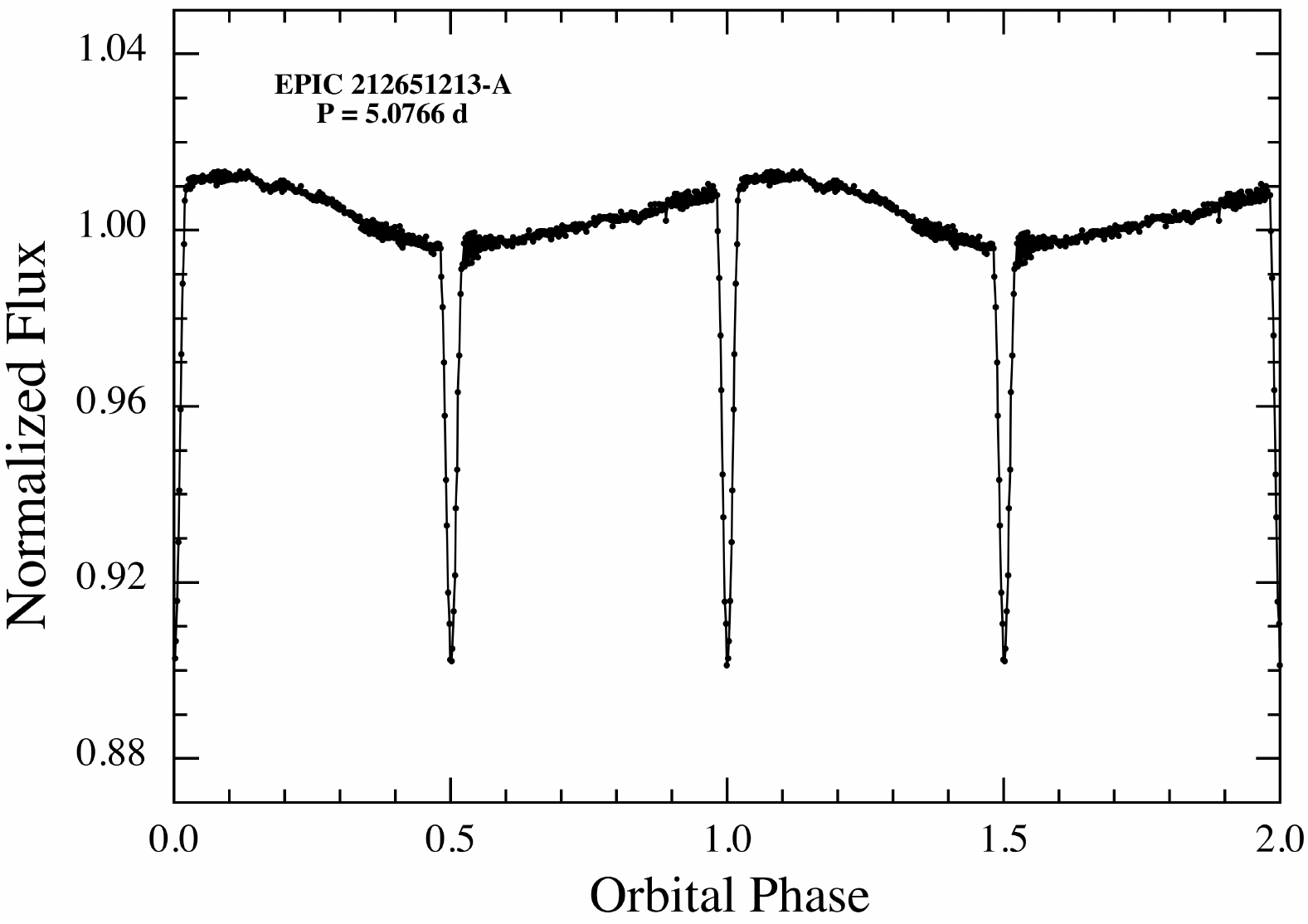} 
\includegraphics[width=0.476 \textwidth]{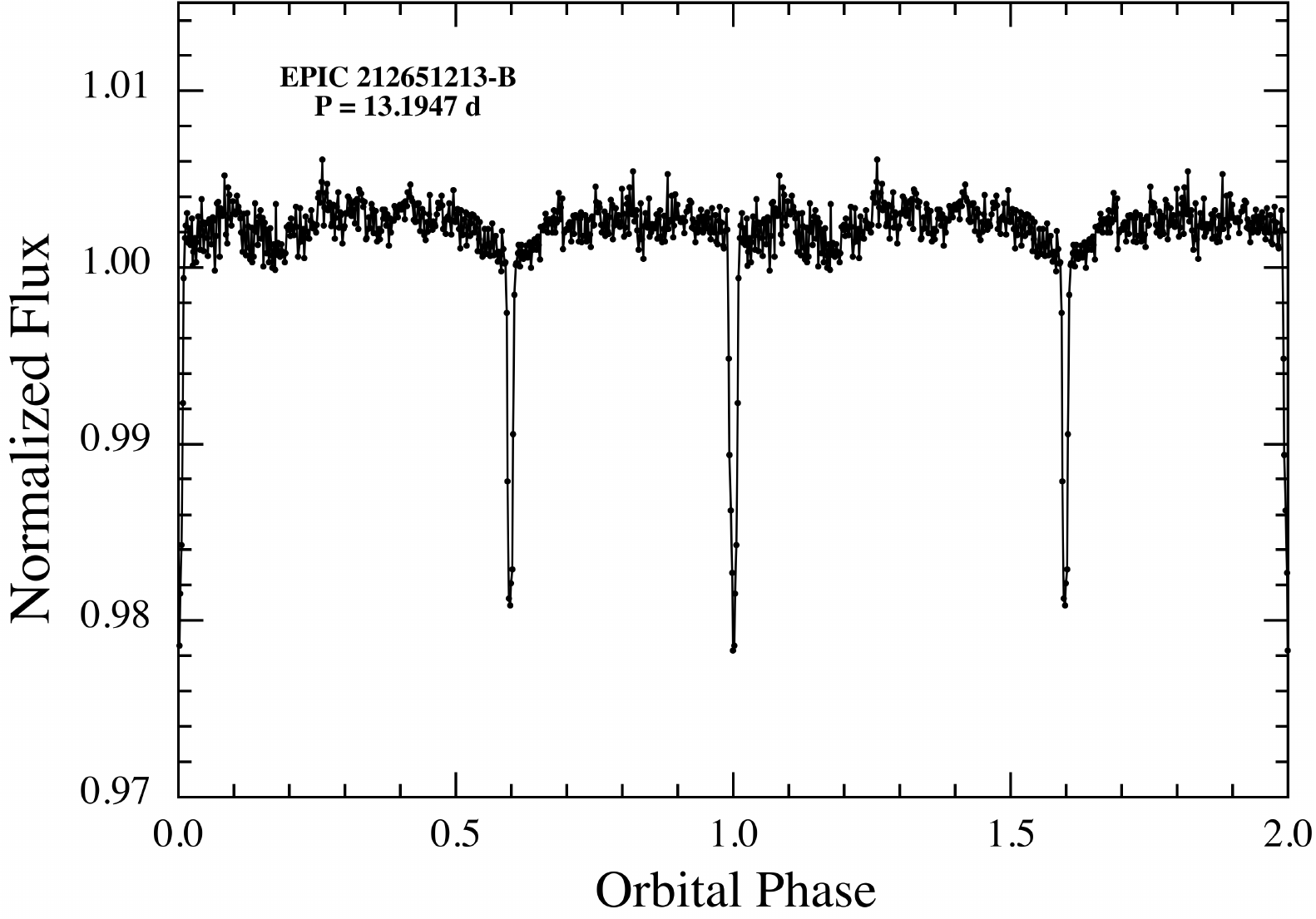} 
\caption{Folded light curves for the 5-day circular binary `A' (top panel) and 13-day eccentric binary `B'.  These have been `disentangled' by subtracting the folded profile of each, sequentially, from the raw flux time series shown in Fig.\ref{fig:rawLC} and then refolding the data.}
\label{fig:folds}
\end{center}
\end{figure}

\begin{table}
\centering
\caption{Radial Velocity Measurements$^a$ of E1213}
\begin{tabular}{lcccc}
\hline
\hline
BJD-2457000 & Star 3 & Star 1 & Star 2 & Observatory$^b$ \\
\hline
437.6168  & +12.19  &	$-81.16$	&  +61.41   & TLS \\
438.6121  &  $-2.51	$ &  $-58.44$  &  +36.12  & TLS \\
439.9122  &  $-16.51$ & +44.81 & $-73.27$ & TRES \\	
446.6095  &  $-37.42$ &	+23.19 & ...  & TLS \\  	  
448.9694  & +31.19	 & $-39.03$ & +31.2 & TRES \\
449.9451  & +27.26	 & ... & $-61.37$ & TRES \\	  	
450.9174  & +10.11	 & 59.15 & $-82.62$ & TRES \\	 	
451.9094  & $-5.34$ & $-5$ & $-5$ & TRES \\	
452.9020  & $-14.53$ & $-81.25$ & +63.81 & TRES \\		
454.8921  & $-35.03$ & +25.78 & $-35.0$ & TRES \\
465.5360  & $-8.09$ & +58.12& $-86.10$ & KO \\
466.4933  & $-22.35$ & ... & ... & KO \\
470.5774	& $-55.28$ & +54.68 & ... & KO \\
470.8631  &  $-53.91$ & +64.01 &$-84.58$ & TRES \\
474.5096  & +11.98 & $-30.42$ & +24.79 & TLS \\
475.5136  & +31.89 & +41.28 & $-78.39$ & TLS \\	
480.4751  & $-27.17$  & +46.06 &  $-71.18$  &  TLS \\
481.4721  & $-36.86$  & +54.41 &  $-80.15$  &  TLS \\
481.5004  & $-37.39$  & +52.06 &  $-78.97$  &  TLS \\
482.4861  & $-44.79$  & $-24.50$ &  +3.67  &  TLS \\
493.4864  & $-25.89$  & $-82.04$ & +64.39 & TLS \\
497.5090  & $-56.17$  & +16.96  & $-0.37$  & TLS \\
498.4503  & $-54.82$  & $-79.09$ & +53.90 & TLS \\
499.4943  & $-34.01$  & $-52.66$ & +40.91 & TLS \\
502.4660  & +25.08     & $-1.47$  & $-26.40$ & TLS \\
502.5105  & ... & +24.23 & $-21.96$ & TLS \\
503.5078  & +8.48 & $-72.66$ & +55.71 & TLS \\
\hline
\end{tabular}
\label{tbl:RVAB}

{\bf Notes.} (a) Units are km s$^{-1}$ and the typical empirical uncertainties on the RV values are 1.7 km s$^{-1}$ for star 3 and 7 km s$^{-1}$ for stars 1 and 2. (b) `TLS' stands for the Th\"uringer Landessternwarte Tautenburg, `KO', the Konkoly Observatory, and `TRES" is the Tillinghast Reflector Echelle Spectrograph of the Whipple Observatory.
\end{table}

\begin{table}
\centering
\caption{Radial Velocity Measurements$^a$ of E1234}
\begin{tabular}{lcc}
\hline
\hline
BJD-2457000 & E1234 & Observatory$^b$\\
\hline
439.9296 & $-14.88$ & TRES \\ 
446.6384 & $-15.63$ & TLS \\
448.9837 & $-14.75$ & TRES \\ 
452.9111 & $-14.73$ & TRES \\ 
465.6044 & $-14.79$ & KO \\
466.9254 & $-14.77$ & TRES \\ 
470.5024 & $-14.59$ & KO \\
475.5430 & $-15.78$ & TLS \\
476.5049 & $-15.53$ & TLS \\
504.8558 & $-14.80$ & TRES \\
511.8791 & $-15.01$ & TRES \\
\hline
\end{tabular}
\label{tbl:RVD}

{\bf Notes.} (a) Units are km s$^{-1}$ and the typical uncertainties on the RV values are 0.1 km s$^{-1}$ for any one observatory. (b) `TLS' stands for Th\"uringer Landessternwarte Tautenburg, `KO', the Konkoly Observatory, and `TRES" is the Tillinghast Reflector Echelle Spectrograph of the Whipple Observatory.
\end{table}

\subsection{RV Analysis}
\label{sec:RVanalysis}

Summaries of the 27 RV observations of E1213 and 11 observations of E1234 are given in Tables \ref{tbl:RVAB} and Table \ref{tbl:RVD}.  In all, the measurements of E1213 span an interval of some 66 days.  

\begin{figure}[h]
\begin{center}
\includegraphics[width=0.476 \textwidth]{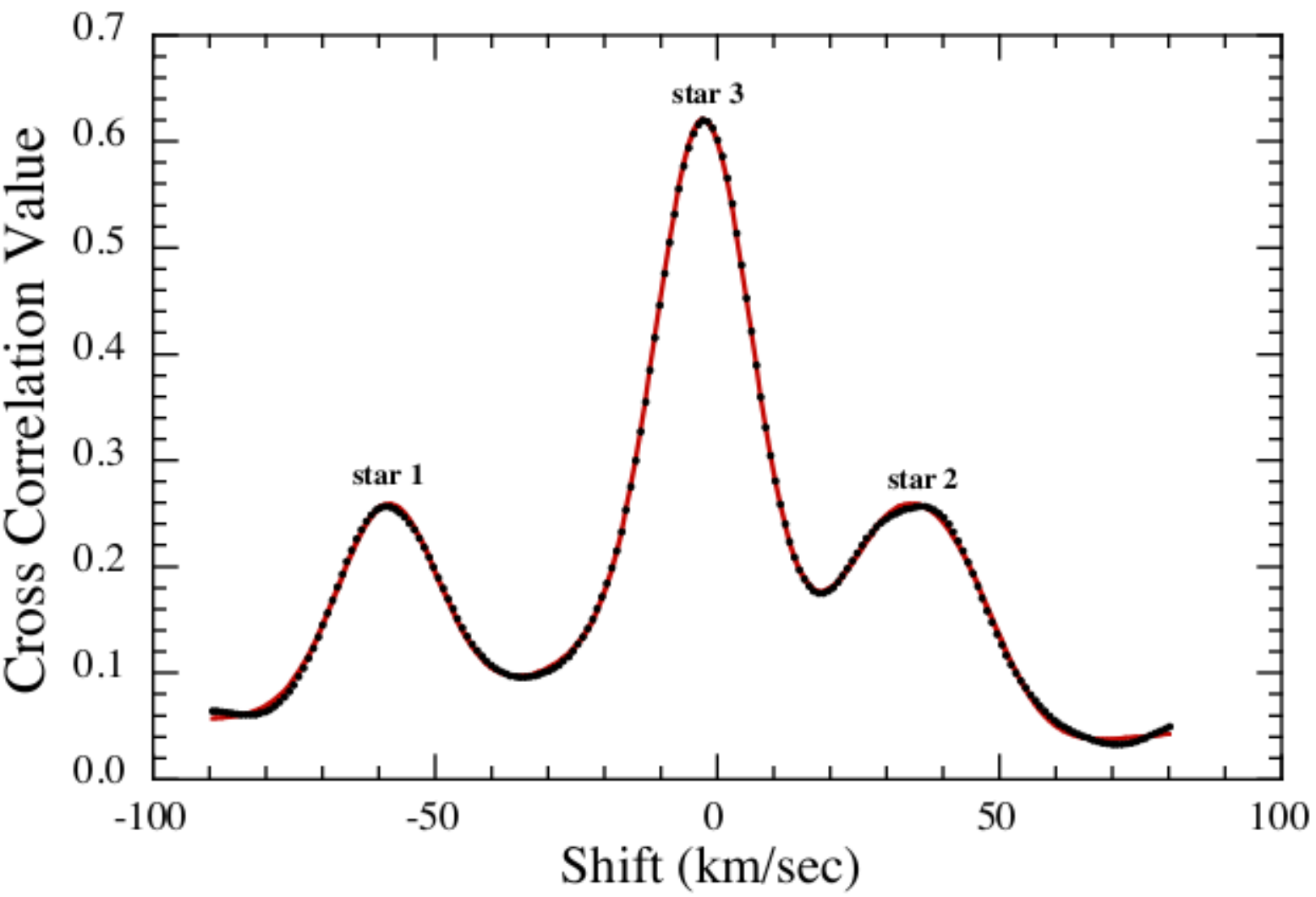} 
\caption{Illustrative cross-correlation function calculated between the observed spectrum and a stellar reference spectrum as a function of the shift in velocities.  The three prominent peaks labeled as `star 1', `star 2', and `star 3', mark the radial velocities of the three stars that can be detected in the CCF.  Stars 1 and 2 comprise the A binary, while star 3 is the brighter of the two stars in the eccentric binary.  The 4th star is not detected in any of the RV spectra.  The red curve is a model fit to the CCF which locates the peaks quantitatively.}
\label{fig:CCF}
\end{center}
\end{figure}

For each observation we cross correlated the acquired spectrum with a single reference template stellar spectrum.  We utilized the template spectrum that yields the highest peak correlation value after testing a grid of $T_{\rm eff}$ and $\log \, g$ values, for an assumed solar metallicity.  An illustrative cross-correlation function (`CCF') is shown in Fig.~\ref{fig:CCF}.  The reference spectrum was for $T_{\rm eff} = 5900$ K and $\log g = 4.2$, close to values ultimately derived for star 3. The three clearly discernible peaks correspond to three stars in the two binaries that are detectable in these RV studies.  The RVs were measured by modeling the CCFs using two profiles with common centers, one Gaussian plus one Cauchy, per stellar component.  When the peaks are clearly separable, as in the example in Fig.~\ref{fig:CCF}, the RVs of these three stars can be well measured.  When the peaks are closer, or even merged, we may be able to determine only the RV of star 3 which dominates the CCF in all observations.  Lines of star 4 could not be detected in any of the spectra.  In particular, we checked when stars 1 and 2 have very small RVs, and simultaneously the RV of star 3 is large---star 4 is still not detected in the CCFs, even with a lower temperature reference spectrum.  The measured RVs for the three detectable stars in E1213 are summarized in Table \ref{tbl:RVAB}.

We also obtained 11 spectra for E1234.  Only a single peak in the CCF is measured in any of these.  The 11 RVs for this image are reported in Table \ref{tbl:RVD}.  We interpret this in terms of either a single star within E1234 or perhaps a binary with an orbit that is sufficiently wide that no RV changes are detected.    We can set a limit of $\lesssim 0.5$ km s$^{-1}$ for RV variations over the 72-day interval of the observations (for more stringent constraints, see Sect.~\ref{sec:E1234}).

\begin{figure}[t]
\begin{center}
\includegraphics[width=0.48 \textwidth]{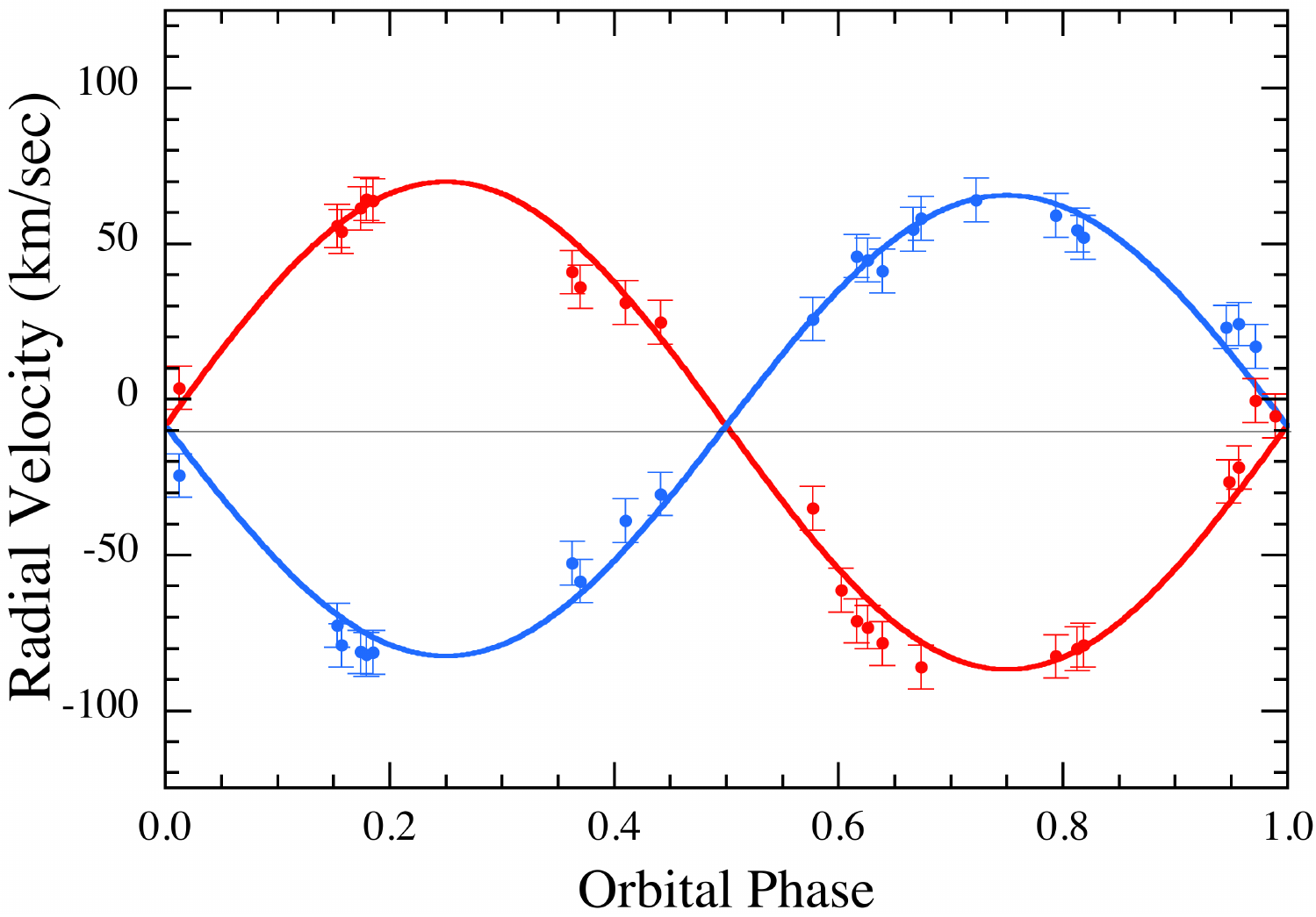}
\includegraphics[width=0.48 \textwidth]{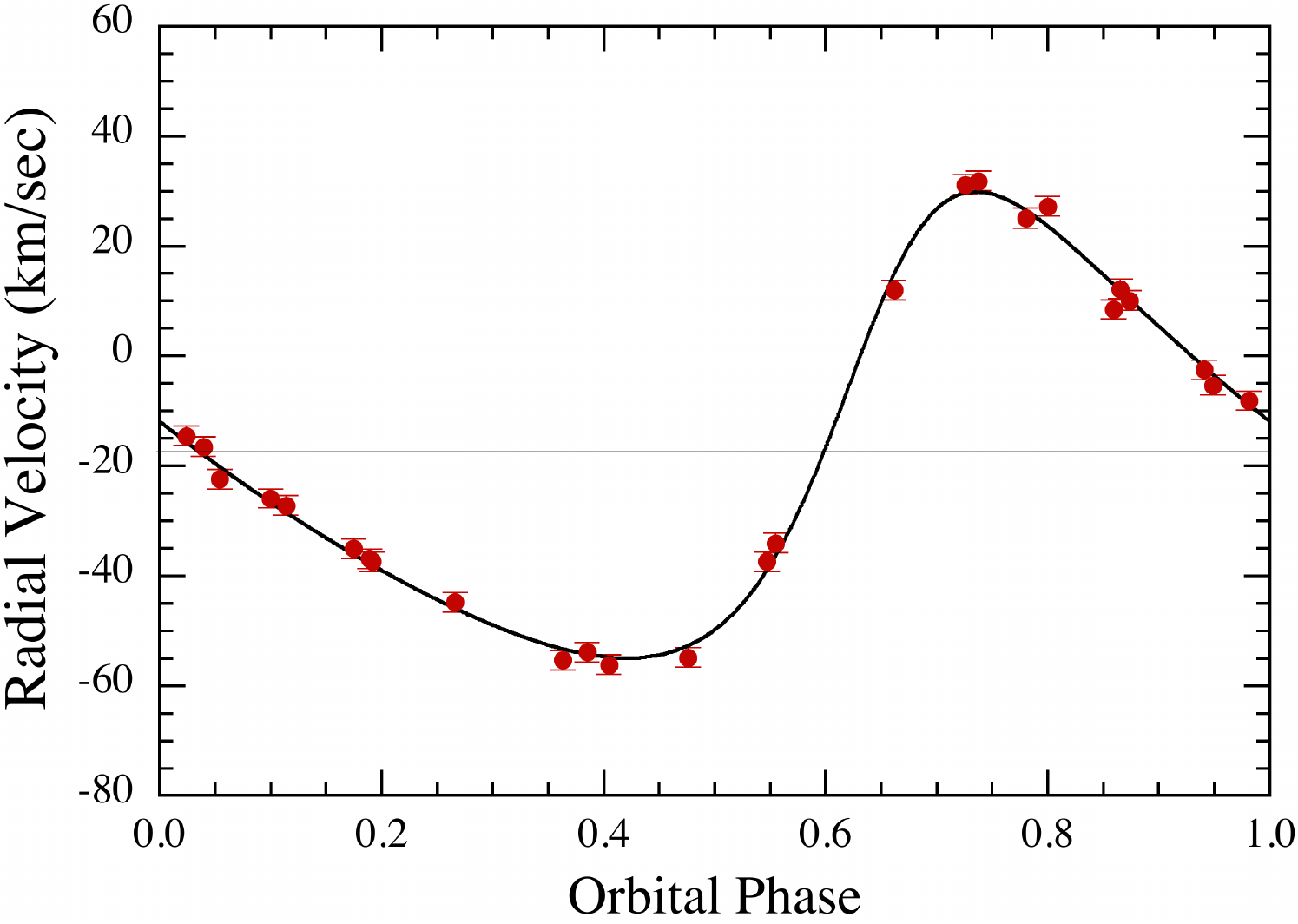}
\caption{Radial velocity curves for the `A' binary (top panel) and the `B' binary (bottom panel).  The solid curves are the best-fitting solutions.  For the `A' binary the fitted parameters are the two $K$ velocities, $\gamma_A$, and orbital phase.  (Blue and red curves are the primary and second star, respectively.) The orbital period was fixed at the value of 5.0766 d as determined from the K2 photometry.  For the `B' binary the fitted parameters are: $K_3$, $e_B$, $\tau_B$, $\omega_B$, and $\gamma_B$ (see Sect.~\ref{sec:RVorbits} for details.)  Note that we have chosen here to define orbital `phase' zero as the time of primary eclipse. }
\label{fig:RVs}
\end{center}
\end{figure}

The RV points for the circular binary, folded modulo the 5-day period are shown in the top panel in Fig.~\ref{fig:RVs}, while those for the 13-day binary are plotted in the bottom panel of Fig.~\ref{fig:RVs}.

\begin{table}
\centering
\caption{Properties of the `A' Binary$^a$}
\begin{tabular}{lcc}
\hline
\hline
Parameter &
Star 1 &
Star 2 \\
\hline
$P_{\rm orb}$ (days)$^b$ & $5.07655 \pm 0.00003$ & $5.07655 \pm 0.00003$ \\
$K$ (km/s)$^c$ &  $73.5 \pm 2.0$ & $78.1 \pm 2.1$ \\  
$\delta K_{\rm rms}$ (km/s)$^c$ & 6.9 & 6.9 \\
$a$ ($R_\odot$)$^c$ &  $7.38 \pm 0.20$  & $7.84 \pm 0.21$ \\ 
$e \cos \omega$$^b$ & $\lesssim 0.0008$ &  $\lesssim 0.0008$ \\
$e \sin \omega$$^b$ & $\lesssim 0.02$  & $\lesssim 0.02$ \\
$e$$^c$ &  $\lesssim 0.02$ & $\lesssim 0.02$ \\ 
$t_{0,\rm{RV}} $ (days)$^c$ & $436.683 \pm 0.025$ & $439.221 \pm 0.025$ \\  
$t_{0,\rm{K2,+35P}} $ (days)$^b$ & $436.634 \pm 0.001$ & $439.172 \pm 0.001$ \\
$t_{0,\rm{K2}} $ (days)$^b$ & $258.9552 \pm 0.0002$ & $261.4935 \pm 0.0002$ \\
$\gamma_A$ (km/s)$^c$ & $-8.4 \pm 1.0$ & $-8.4 \pm 1.0$ \\
$\dot \gamma_A$ (cm s$^{-2}$)$^c$ & $< 0.01$ & $<0.01$ \\
$f(M_1)$ ($M_\odot$)$^d$ &  $0.210 \pm 0.017$ & $0.252 \pm 0.020$ \\
$i_A$$^e$ & $85.7^\circ \pm 0.5^\circ$ & $85.7^\circ \pm 0.5^\circ$ \\
$M$ $(M_\odot)$$^f$ & $0.94 \pm 0.06$ & $0.89 \pm 0.05$ \\
$R$ $(R_\odot)$$^g$  & $0.86 \pm 0.12$ & $0.80 \pm 0.09$ \\
$T_{\rm eff}$ (K)$^g$ & $5475 \pm 200$ & $5250 \pm 200$ \\
$\log$ g (cgs)$^g$ & $4.54 \pm 0.10$ & $4.59 \pm 0.07$ \\
\hline
\end{tabular}
\label{tbl:Abinary}

{\bf Notes.} (a) All values are cited with 1-$\sigma$ uncertainties; times are BJD-2457000. (b) Based on the K2 photometry. (c) From fits to the radial velocity data given in Table \ref{tbl:RVAB}, with eccentricity fixed at $e = 0$. (d) Mass function for each star based on their respective $K$ velocities. (e) From an analysis using the Phoebe binary light curve emulator (Pr\v{s}a \& Zwitter 2005). (f) From a Monte Carlo analysis of the system parameters and error propagation (see Sect.~\ref{sec:MC_eval}).  (g) The extra constraints needed to yield the full set of system parameters were obtained from observational restrictions on the light ratio between the `A' and `B' binaries, and the Yonsei-Yale evolution tracks (see Sect.~\ref{sec:MC_eval}).
\end{table}

\section{Radial Velocity Orbits}
\label{sec:RVorbits}

\subsection{Fitting the `A' binary}

Given the very small value of $e \lesssim 0.02$ for the `A' binary that we were able to deduce from the K2 light curve (see Sect.~\ref{sec:K2}), we restricted our fits to circular orbits.  We fit the RV points shown in the upper panel of Fig.~\ref{fig:RVs} simultaneously for the orbits of stars 1 and 2 using common values of the orbital phase and $\gamma_A$ velocity, as well as $K_1$ and $K_2$ velocities.  We fixed the orbital period at $P_{\rm orb} = 5.0766$ d, as determined from the K2 data.  This is more accurate than the period we would be able to determine from the RV data alone.  We utilized a simple Levenberg-Marquardt fitting routine to determine the four unknown parameters ($K_1$, $K_2$, $\gamma_A$, $t_{\rm 0,RV}$) and their uncertainties.  

The best-fit model RV curves for the `A' binary are shown in the top panel in Fig.~\ref{fig:RVs}, and the fitted orbital parameters are summarized in Table \ref{tbl:Abinary}.  The values of $K_1$ and $K_2$ are 73.5 km s$^{-1}$ and 78.1 km s$^{-1}$, respectively; these are similar but do differ at the 1.6-$\sigma$ level.  The orbital phase agrees with the phase obtained from the K2 photometry and projected forward by nearly half a year.  The agreement is good to within 0.047 days with an uncertainty of 0.025 days (i.e., $\sim$$1.1 \pm 0.6$ hr).  
	
We also note that star 1, with the somewhat lower $K$ velocity (compared to star 2), and hence the higher mass, is the one whose descending node on the RV curve is an integer number of orbital cycles from the slightly deeper eclipse in the K2 light curve (see the top panel of Fig.~\ref{fig:folds}).  This is the time when the more massive star is eclipsed.  And, at least for stars on the main sequence, the more massive star should have the higher surface brightness, and hence the deeper eclipse when it is transited---just as we observe.

We use the $K$ values to derive the constituent masses in Sect.~\ref{sec:MC_eval}.

\begin{table}
\centering
\caption{Properties of the `B' Binary$^a$}
\begin{tabular}{lc}
\hline
\hline
Parameter &
Value 
 \\
\hline
$P_{\rm orb}$ (days)$^b$ & $13.1947 \pm 0.0004$  \\
$K_3$ (km/s)$^c$ &  $42.57 \pm 0.28$   \\ 
$\delta K_{\rm rms}$ (km/s)$^c$ & 1.77  \\
$a_3$ $ (R_\odot$)$^c$ &  $10.51 \pm 0.07$   \\
$e$$^c$ &  $0.325 \pm 0.006$  \\ 
$\omega_B$ (deg)$^c$ & $298.7 \pm 1.2$ \\
$\tau_B$ (days)$^c$ & $434.73 \pm 0.05$  \\   
$t_{\rm ecl,RV}$ (days)$^c$ & $439.45 \pm 0.05$ \\ 
$t_{\rm ecl, K2+16P}$ (days)$^b$ & $439.39 \pm 0.01$ \\ 
$t_{\rm ecl, K2}$ (days)$^b$ & $228.270 \pm 0.003$ \\
$\gamma_B$ (km/s)$^c$ & $-19.0 \pm 0.2$  \\
$\left| \dot \gamma_B \right|$ (cm s$^{-2}$)$^c$ & $ < 0.007$ \\
$f(M_3)$ ($M_\odot$)$^d$ &  $0.089 \pm 0.002$   \\
$i_B$$^e$ & $85.8^\circ \pm 0.1^\circ$  \\
$M_3$ $(M_\odot)$$^f$ & $1.09 \pm 0.07$  \\
$M_4$ $(M_\odot)$$^f$ & $0.64 \pm 0.03$  \\
$R_3$ $(R_\odot)$$^f$ & $1.07^{+0.17}_{-0.11}$ \\
$R_4$ $(R_\odot)$$^f$ & $0.57 \pm 0.03$ \\
$T_{\rm eff,3}$ (K)$^f$ & $6000 \pm 280$ \\
$T_{\rm eff,4}$ (K)$^f$ & $4280 \pm 110$ \\
$\log$ g$_3$ (cgs)$^f$ & $4.43^{+0.10}_{-0.16}$\\
$\log$ g$_4$ (cgs)$^f$ & $4.75 \pm 0.03$ \\
\hline
\end{tabular}
\label{tbl:Bbinary}

{\bf Notes.} (a) All values are cited with 1-$\sigma$ uncertainties; times are BJD-2457000. (b) Based on the K2 photometry. (c) From fits to the radial velocity data given in Table \ref{tbl:RVAB}. (d) Mass function based on the $K$ velocity and the orbital eccentricity. (e) From an analysis using the Phoebe binary light curve emulator (Pr\v{s}a \& Zwitter 2005). (f) The extra constraints needed to yield the full set of system parameters were obtained from observational restrictions on the light ratio between the `A' and `B' binaries, and the Yonsei-Yale evolution tracks (see Sect.~\ref{sec:MC_eval}).  
\end{table}

\subsection{Fitting the `B' binary}

We next fit the RV points plotted in the bottom panel of Fig.~\ref{fig:RVs} for the obviously eccentric orbit.  We held the orbital period fixed at 13.1947 days, as determined from the K2 photometry, and fit for 5 free parameters: $a_3 \, \sin i_B, \, e_B, \, \omega_B, \tau_B, \, {\rm and} \,  \gamma_B$, where $\omega_B$ and $\tau_B$ are the longitude and passage time of periastron.  For this we utilized a Markov chain Monte Carlo (`MCMC') code to map out the uncertainties in the parameters.  The results are reported in Table \ref{tbl:Bbinary}.  In spite of the fact that this is a single-line spectroscopic binary, we were still able to make useful mass estimates of both star 3 and star 4 via their combined flux compared to the flux from binary ``A'' (see Sect.~\ref{sec:MC_eval}).

\section{Evaluation of Basic Stellar and Atmospheric Parameters}
\label{sec:MC_eval}

We now have two independent mass functions for star 1 and star 2 in the `A' binary, and a mass function for star 3 in the `B' binary.  For both binaries, the orbital inclination angles are determined with the {\tt Phoebe} binary light curve emulator (Pr\v{s}a \& Zwitter 2005), and are very close to 86$^\circ$.
To determine the masses and other stellar properties we utilize two ingredients: (i) the three mass functions and their uncertainties (Tables \ref{tbl:Abinary} and \ref{tbl:Bbinary}), and (ii)  the relative fluxes of the ``A'' and ``B'' binaries (see, e.g., Sect.~\ref{sec:speckle}).  For the latter constraint we adopt the following specific equality involving the stellar luminosities:
\begin{eqnarray}
\frac{(L_3+L_4)}{(L_1+L_2)} = 1.6 \pm 0.16 ~~(1\sigma)  
\label{eqn:lum}
\end{eqnarray}

We then use a Monte Carlo error propagation technique to evaluate all four masses simultaneously.  First we choose realizations for stars 1 and 2, using the cited values for the orbital $K$ velocities and random Gaussian draws of their cited uncertainties (Tables \ref{tbl:Abinary} and \ref{tbl:Bbinary}).  This then determines $M_1$ and $M_2$ for that particular realization of $K_1$ and $K_2$.  Next we similarly chose a random realization of the value of $K_3$ using its error bar.  We then choose a random mass ratio for $M_4/M_3$ between 0 and 1; that, plus the mass function for star 3, yields a specific realization for $M_3$ and $M_4$. 

\begin{figure}[b]
\begin{center}
\includegraphics[width=0.48 \textwidth]{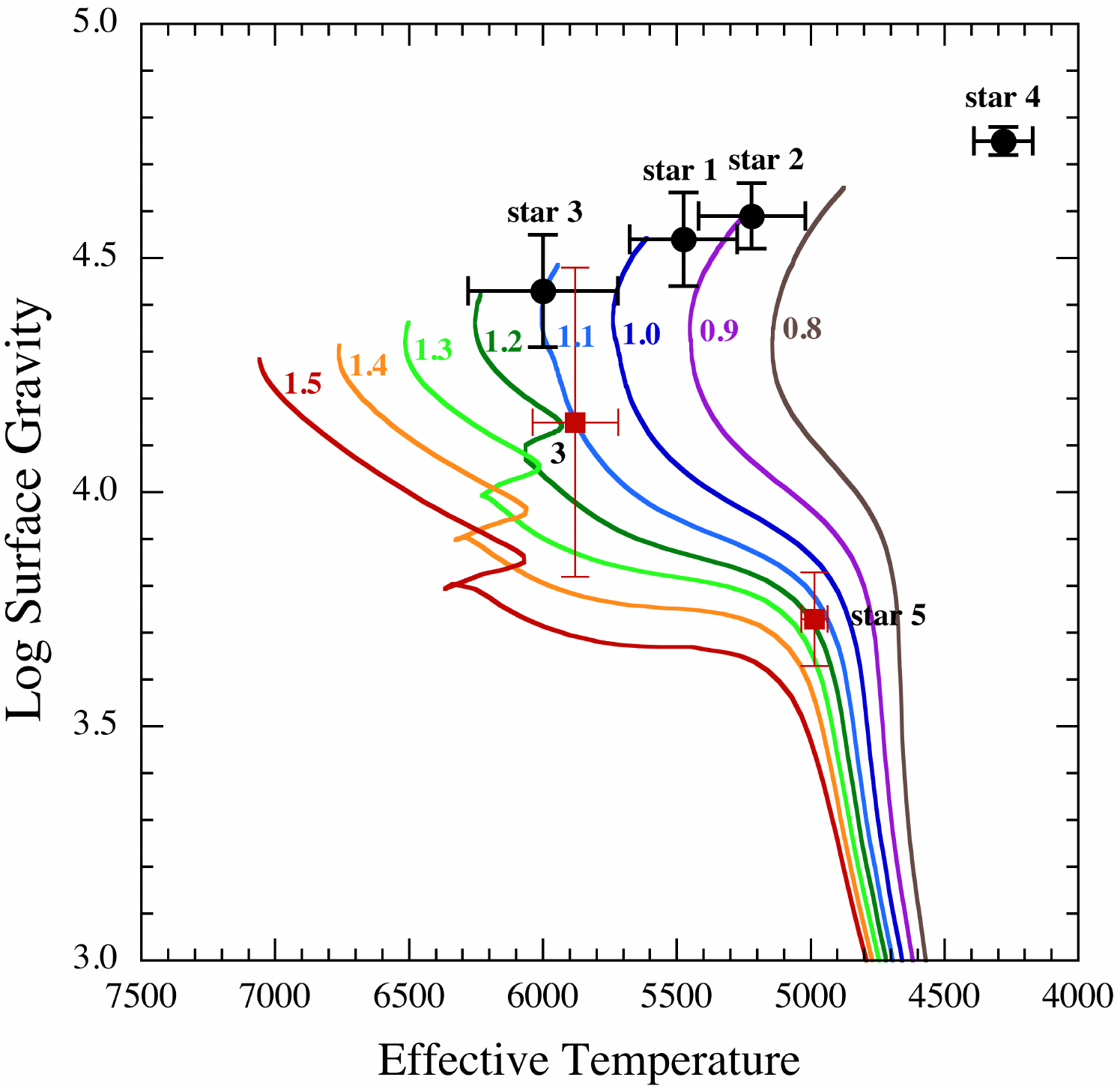}
\caption{Evolution tracks for stars of mass 0.8 to 1.5 $M_\odot$ in the $\log \, g - T_{\rm eff}$ plane.  These were computed from the Yonsei-Yale tracks (Yi et al.~2001) for Z = 0.02.  The black circles with error bars are the locations of stars 3, 1, 2, and 4 from left to right respectively, as determined from our Monte Carlo analysis of the system parameters (see Sect.~\ref{sec:MC_eval}).  The red error bars for star 3 show for comparison its location based on the spectrum analysis (see Sect.~\ref{sec:decomp}).  Stars 1 and 2 comprise binary ``A'', stars 3 and 4 are in binary ``B'', while star 5 is the E1234 image. The location of star 5 is determined entirely from the spectral analysis study and is also plotted in red (see Sect.~\ref{sec:decomp}).}
\label{fig:YY}
\end{center}
\end{figure}

In order to further constrain the system parameters, we require at least one more substantive constraint on the system.  We take this supplemental information to be the measured ratio of the brightness of binary `B' to binary `A' (see Sect.~\ref{sec:AO}).  However, in order to incorporate the luminosities of the star, we need to make use of basic evolution tracks for low-mass stars.  In order to do that, we have devised non-physical fitting functions to the Yonsei-Yale evolution tracks (see Fig.~\ref{fig:YY}; Yi et al.~2001).  The following represent the radius and luminosity evolution of a star of mass $m$ and evolution age, $t$.  
\begin{eqnarray}
R(m,t) \simeq 0.84\,m \,e^{m^{3.5}t/26} \left[1+e^{(t-12.5m^{-3.5})/0.9}\right] \\
L(m,t) \simeq 0.66\,m^{4.6} \,e^{m^{3.5}t/12.3} \left[1+e^{(t-12.5m^{-3.5})/0.7}\right] 
\label{eqn:fitting}
\end{eqnarray}
where $m$ is in solar masses, and the evolution age, $t$, is expressed in Gyr.  These should be sufficiently accurate for stars of mass $0.8 \lesssim m \lesssim 1.4$.  

Thus, our Monte Carlo realizations of the system also require an evolutionary age, which we take to be uniformly distributed between 1 and 10 Gyr.  Once we have a set of 4 masses and a randomly chosen evolution age, we can use Eqns.~(4) and (\ref{eqn:fitting}) to evaluate all the system parameters, including masses, radii, luminosities, and values of $\log$ g.  The ratio of $(L_3+L_4)/(L_1+L_2)$ is then computed for the given realization.  This ratio is then checked against the measured ratio given in Eqn.~(\ref{eqn:lum}) with a randomly chosen value with respect to the uncertainty given in that equation.  If the computed ratio lies outside that of the `measurement', the entire system is discarded.

\begin{figure*}
\begin{center}
\includegraphics[width=0.48 \textwidth]{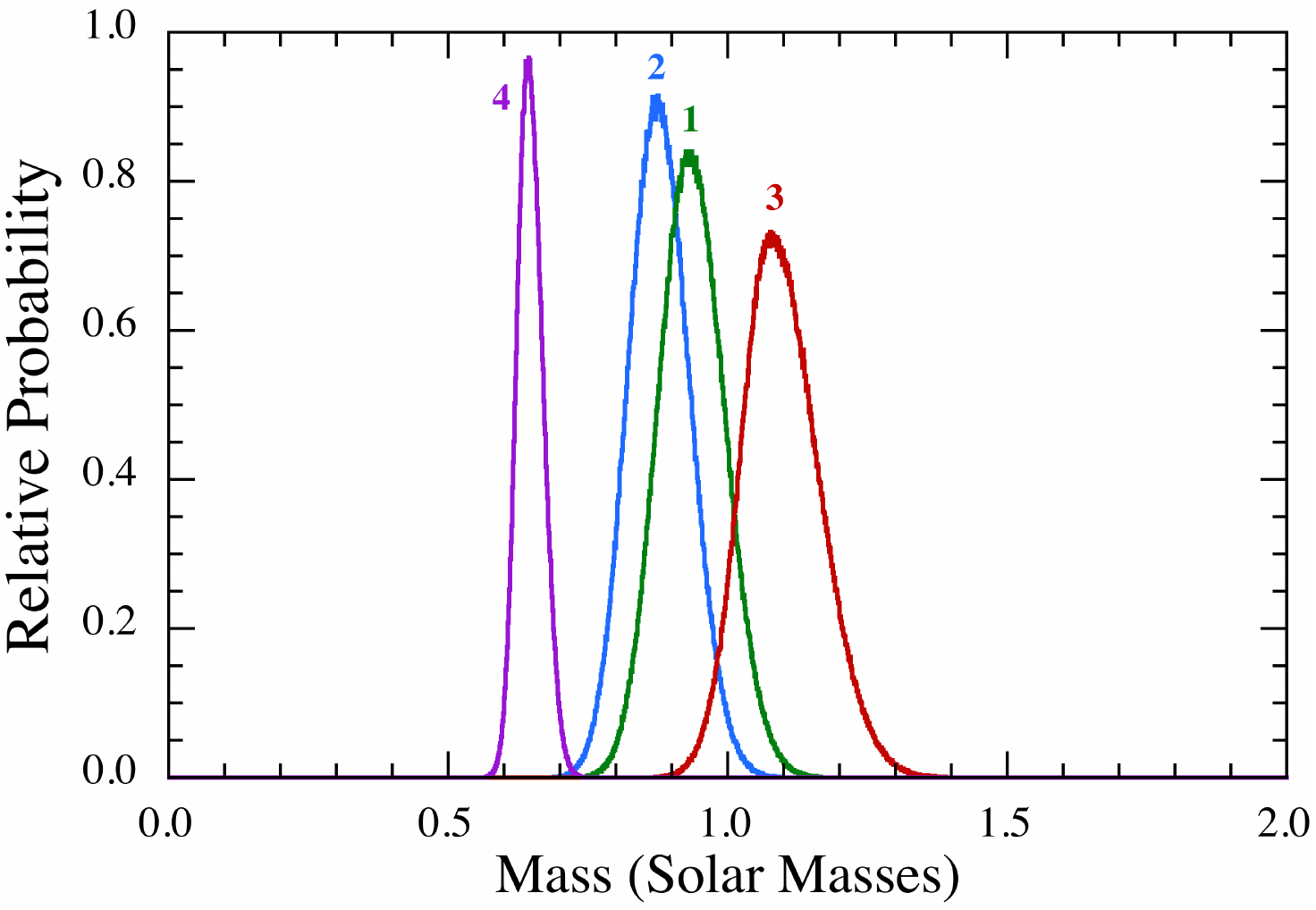} \hglue0.1cm
\includegraphics[width=0.48 \textwidth]{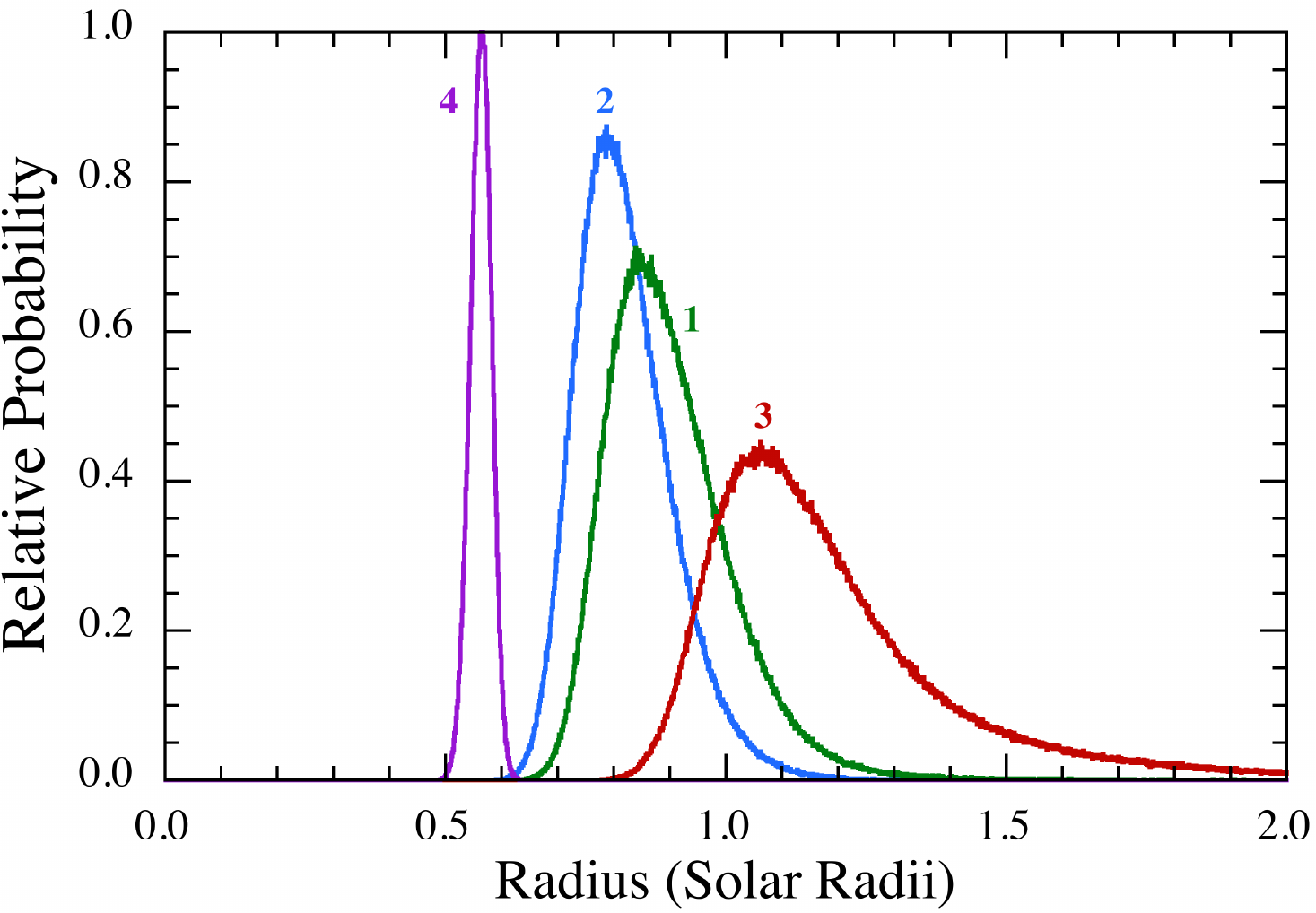} \vglue0.1cm
\includegraphics[width=0.48 \textwidth]{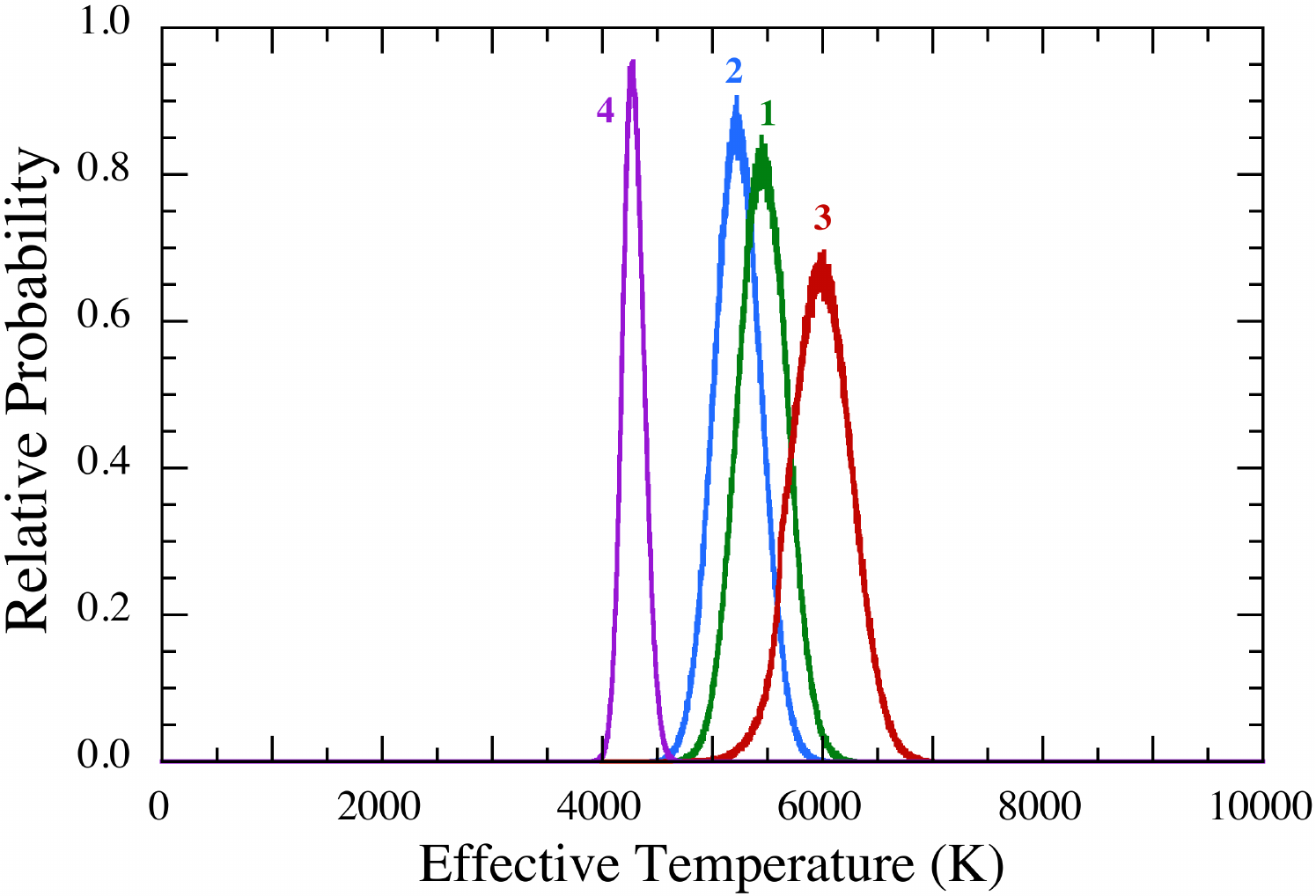} \hglue0.1cm
\includegraphics[width=0.48 \textwidth]{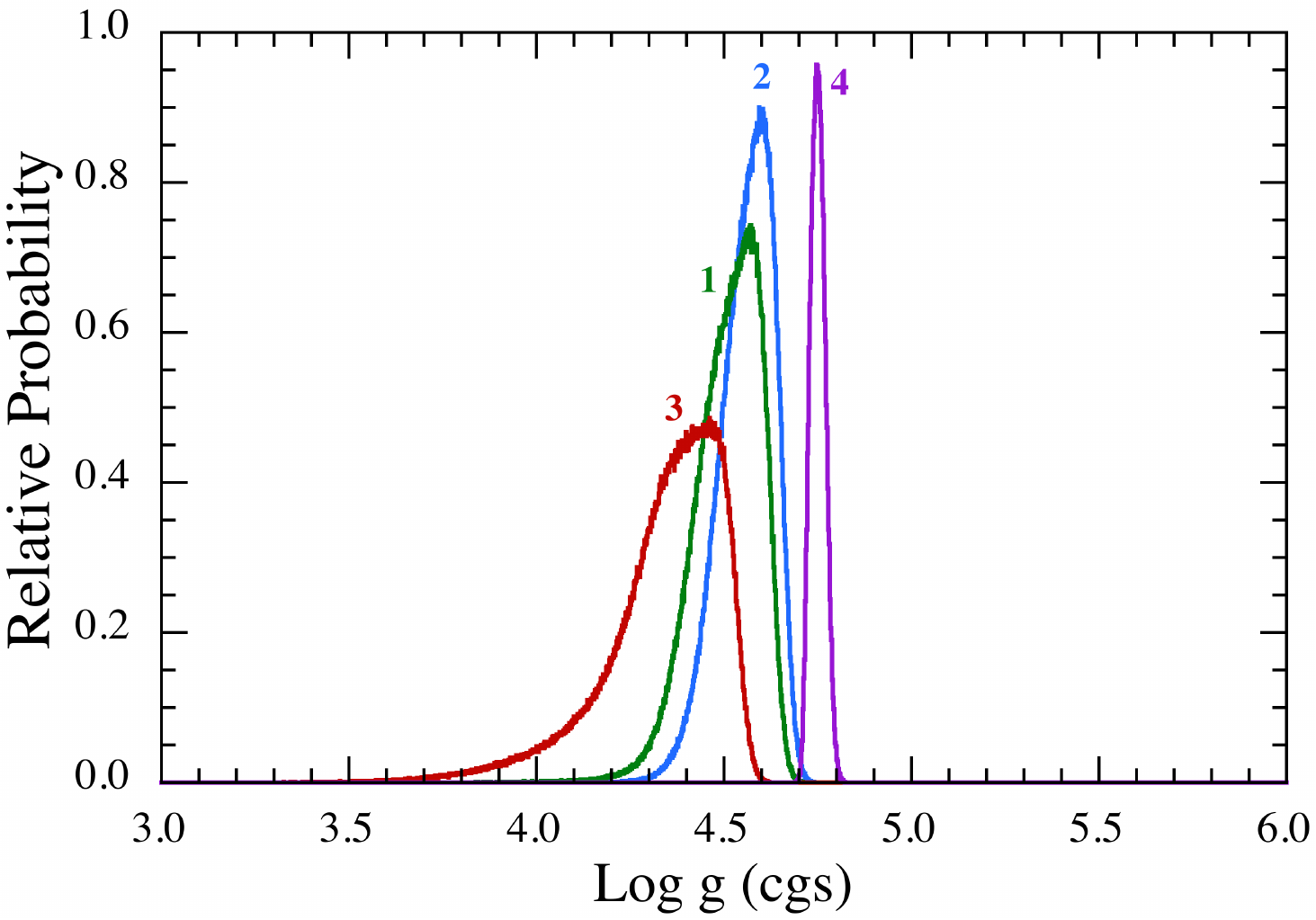}
\caption{Probability distributions for the masses, radii, effective temperatures, and values of $\log \, g$ of the four stars in binaries `A' and `B'.  These were computed from a Monte Carlo error propagation code using the uncertainties in the three $K$ velocities.  To close the equations, we also added the constraint from the speckle images (see Sect.~\ref{sec:speckle}) that the luminosity of binary ``B'' exceeds that of binary ``A'' by a factor of $1.6 \pm 0.16$ (see Eqn.~\ref{eqn:lum}). }
\label{fig:dist}
\end{center}
\end{figure*}

This process is repeated $10^8$ times, and distributions for the four masses, radii, $T_{\rm eff's}$, and $\log$ g's are built up.  The results are shown in Fig.~\ref{fig:dist}.  A perusal of these mass distributions reveals a few interesting things.  For the `A' binary, the masses of stars 1 and 2 are: $0.94 \pm 0.06 \, M_\odot$ and $0.89 \pm 0.05 \, M_\odot$, respectively.  The most probable values of $\log$ g are 4.54 and 4.59, with an uncertainty of 0.1 dex, respectively.  The locations of stars 1 and 2 in the $\log \, g - T_{\rm eff}$ plane are shown in Fig.~\ref{fig:YY}, and they are clearly in good accord with the corresponding mass of the tracks they are near.   

As for the `B' binary, the mass distributions for star 3 and star 4 (see Fig.~\ref{fig:dist}) indicate $M_3 \simeq 1.09 \pm 0.07$ and $M_4 \simeq 0.64 \pm 0.03$.  These are rather well defined given that this is a single-line spectroscopic binary, and result from the luminosity comparison of binary ``B'' with binary ``A''.  The most probable values for $\log$ g are $4.43^{+0.10}_{-0.16}$ and $4.75 \pm 0.03$, respectively.  The locations of these stars comprising binary ``B'' are shown in the $\log \, g - T_{\rm eff}$ plane in Fig.~\ref{fig:YY}.  Star 3 appears to be slightly evolved.

\section{Spectral analysis}
\label{sec:decomp}

The TLS spectra were taken during grey time and some of them are slightly 
contaminated with moonlight. Though this contamination could be separated from 
the stellar features in most of the CCFs used for the RV determination, we had 
to be sure that the contamination from the moonlight does not enter the spectrum analysis. 
We therefore initially used only nine of the 16 TLS spectra of E1213 listed in Table 2 for the analysis. 
This number was too small for a secure decomposition of the observed composite spectra into 
the spectra of the individual stellar components. Nonetheless, we tried the FFT-based KOREL 
program (Hadrava 1995; 2004) that delivers the orbital parameters together with the RVs 
and the decomposed spectra, and then the singular value decomposition of 
Simon \& Sturm (1994) with the measured RVs as input values using
the CRES code (Ilijic 2004).  However, in both cases the 
subsequent analysis of the decomposed spectra gave inconsistent results. In the end, 
we decided to analyze only the one composite spectrum that shows the largest 
separation of the components' peaks in the CCF (Fig.~\ref{fig:CCF}).

\begin{figure}
\begin{center}
\includegraphics[width=0.48 \textwidth]{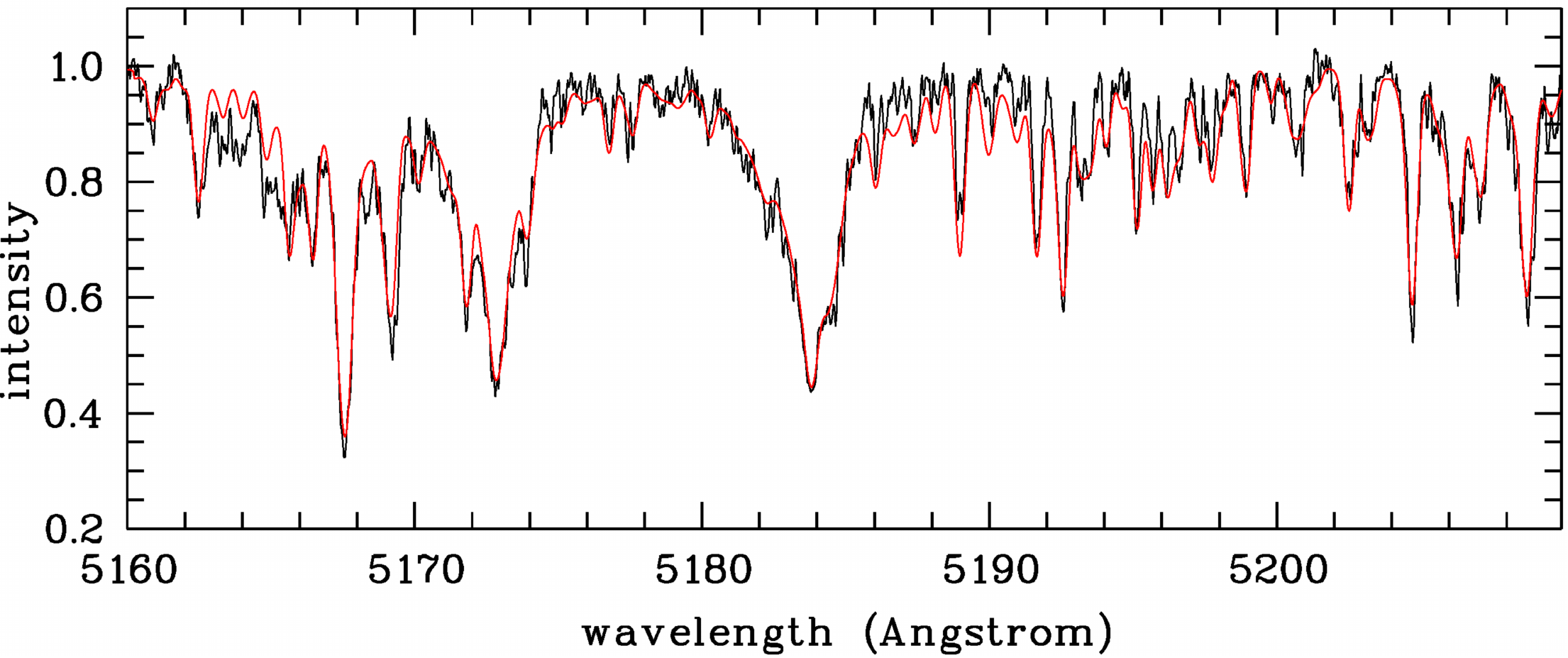}
\caption{Part of the analysed composite spectrum of E1213 (black color) showing the Mg I triplet region, together with the best fitting combination of synthetic spectra (red color; see Table \ref{tbl:decomp}, Stars 1 to 3). }
\label{fig:Mgtriplet}
\end{center}
\end{figure}

Spectrum analysis was performed using the GSSP 
program (see Lehmann et al.~2011 and Tkachenko et al.~2012 for a 
description of the method). The program is based on the spectrum synthesis 
approach. We extended the method to a simultaneous fit for three components in a
composite spectrum. The synthetic spectra were computed on the TLS cluster
computer with the parallelized version of the SynthV program (Tsymbal 1996), 
using a library of precomputed line-by-line model atmospheres (Shulyak et al.~2004). 
The atomic data were taken from the VALD data base (Kupka et al.~1999). 
We used three grids of atmospheric parameters, one for each stellar component. 
Free parameters were [M/H], $T_{\rm eff}$, $\log \,g$, microturbulent velocity
$v_{\rm turb}$, and $v\sin i$. For the surface metalicities [M/H] we used
scaled solar abundances in accordance with Asplund et al.~(2009). Moreover, we did 
a fine-adjustment for the RVs of the contributions of the three components, using the 
values obtained from fitting the CCF of the spectrum (Sect.~\ref{sec:RVanalysis}) as the starting values. 
Normally, when we deal with an SB3 star, the flux ratios of the three stars
compared to the total continuum flux are coupled via $f_1+f_2+f_3=1$ and can be
determined for each grid point from least squares minimisation of two of the
flux ratios. In our case, we got a much better fit to the observed composite spectrum 
(see Fig.~\ref{fig:Mgtriplet} with the fit of the Mg I triplet) when adding an 
additional continuum contribution (veiling), originating from star 4. In that 
case, $f_4$ is determined from $1-(f_1+f_2+f_3)$ and $f_1$ to $f_3$ are 
no longer coupled via their flux ratios and were treated as free parameters of the fit.

\begin{table}
\centering
\caption{Results from Spectrum Analysis}
\begin{tabular}{lcccc}
\hline
\hline
Parameter &
Star 1 & 
Star 2 & 
Star 3 & 
Star 5$^a$
 \\
\hline
$T_{\rm eff}$ (K)  & $5700 \pm 250$ & $5490 \pm 310$ & $5880 \pm 160$ & $4987 \pm 50$  \\ 
$\log \, g$ (cgs) &  $\{4.5\}^b$ & $\{4.6\}^b$ & $4.15 \pm 0.33$ & $3.73 \pm 0.1$ \\ 
~[M/H]  & $\{0.1\}^b$ & $\{0.1\}^b$ & $0.08 \pm 0.20$ & $0.34 \pm 0.08$ \\ 
$v_{\rm turb}$ (km s$^{-1}$)  & $1.6 \pm 0.6$ & $1.7 \pm 0.7$ & $1.6 \pm 0.4$ &  \{1.0\}$^b$\\
$v_{\rm sin\,i}$ (km s$^{-1}$)  & $13 \pm 5$ & $7 \pm 6$ & $14 \pm 1.8$ & $2.7 \pm 0.5$ \\ 
$f$$^c$  & $0.20 \pm 0.05$ & $0.13 \pm 0.04$ & $0.59 \pm 0.05$ & ... \\
\hline
\end{tabular}
\label{tbl:decomp}

{\bf Notes.} (a) Star 5 refers to the image of E1234.  (b) Quantities in curly brackets were held fixed at plausible representative values. (c) Fraction of the light in stars 1--4.  
\end{table}

We used all five of the above mentioned quantities as free parameters for star 3. The relatively 
low S/N of $\sim$32 for the single spectrum, together with the low line strengths of stars 1 
and 2 compared to star 3, did not allow us to do the same for stars 1 and 2. For these, we 
fixed $\log \, g$ at 4.5 and 4.6, respectively, as the most probable values obtained in Sect.~\ref{sec:MC_eval},
and [M/H] at the value obtained for star 3 (the latter was done iteratively). 
After the final optimal solution for all free parameters was found, the parameter 
errors were determined from $\chi^2$ statistics using the full grid of atmospheric 
parameters. In this way, the errors include all interdependencies among the 
parameters. 

A comparison of $\log \, g$ and $T_{\rm eff}$ for star 3 derived from this spectrum analysis (Table \ref{tbl:decomp})
with those inferred from the Monte Carlo evaluation of the system parameters (see Sect.~\ref{sec:MC_eval})
is shown in Fig.~\ref{fig:YY}.  The values of $\log \,g$ differ by 0.28 dex between that found from the spectrum
analysis vs.~that from the MC analysis.  While this difference is only about 1-$\sigma$, it is nonetheless suggestive that star 3
may be somewhat more evolved than the MC analysis indicates.  
The inferred values of the flux ratio of binary `B' to binary `A' from the spectrum analysis is $2.03 \pm 0.45$ which
is consistent with the value of $1.58 \pm 0.16$ derived from the speckle imaging (see Sect.~\ref{sec:speckle}).  
Overall, we take the spectrum analysis to be a satisfactory check on our more detailed parameter study in Sect.~\ref{sec:MC_eval}.

Finally, we obtained the stellar parameters for E1234 using the Spectral Parameter Classification (SPC) tool developed by Buchhave et al.~(2012). SPC cross correlates an observed spectrum against a grid of synthetic spectra based on Kurucz atmospheric models (Kurucz 1992). The weighted average for the six TRES spectra of E1234 is $T_{\rm eff}=4987\pm 50$ K, $\log \, g = 3.73 \pm 0.10$, [m/H] = $0.34 \pm 0.08$, $v\, \sin\,i=2.7 \pm 0.5$ km s$^{-1}$. The values were calculated by taking an average of the stellar parameters that were calculated for each spectrum individually. The values are then weighed according to the cross correlation function peak height.  

Table \ref{tbl:decomp} summarizes the spectral analysis results for both E1213 and E1234, while Fig.~\ref{fig:YY} shows where the different stars fall in the $\log \, g - T_{\rm eff}$ plane.

\section{Tidal Synchronization Status of `A' and `B'}
\label{sec:tides}

The spectroscopically obtained projected rotational velocities of stars 1--3 (see Table \ref{tbl:decomp}) carry indirect information on
the state of the tidal synchronization processes in the two eclipsing binaries. The
projected synchronized (or, in the case of an eccentric system: pseudo-synchronized) rotational velocity of the binary members can be calculated simply as
\begin{equation}
\left(v_{\rm star}\sin{i_\mathrm{orb}}\right)_\mathrm{sync}=\frac{2\pi R_{\rm star}\sin{i_\mathrm{orb}}}{P_\mathrm{orb}\mathcal{P}(e)},
\end{equation}
where $\mathcal{P}(e)$ the orbital eccentricity-dependent ratio of the orbital to the (pseudo-)synchronized rotational
period which, in the equilibrium tide model of Hut (1981) can be calculated as
\begin{equation}
\mathcal{P}(e)=\frac{\left(1-e^2\right)^{3/2}\left(1+3e^2+\frac{3}{8}e^4\right)}{1+\frac{15}{2}e^2+\frac{45}{8}e^4+\frac{5}{16}e^6} ~< 1.
\end{equation}
Furthermore, we also assumed that for (pseudo-)synchronized rotation the orbital angular momentum vectors of the star and orbit are aligned.  Substituting stellar radii, orbital inclinations, periods, and eccentricities from Tables \ref{tbl:Abinary} and \ref{tbl:Bbinary} we arrive at
$(v_{\rm star}\sin{i})_{\rm obs}/(v_{\rm star}\sin{i})_{\rm sync}=\,1.5 \pm 0.6$, $0.9 \pm 0.8$, and $2.1 \pm 0.4$ for stars 1, 2 and 3, respectively.
Therefore, we can conclude that the stars in the 5-day circularized binary `A', are rotationally synchronized with a significant likelihood.
(Although, theoretically, we cannot exclude the possibility of faster stellar rotations around misaligned rotational axes.)  For the primary star
of the eccentric binary `B', our result strongly supports a modestly super-synchronous rotation. 

According to the formulae of Moreno et al.~(2011) and Zahn (2008), for a convective star in an eccentric $e\sim0.3$
orbit with a fractional radius of $R_{\rm star}/a\sim0.04$, the expected synchronization time-scale even for an initially
ten-times faster orbital rotation should be $\lesssim $ 1 Gyr for physically realistic values of the viscosity parameter. 
This timescale is shorter than the likely age we infer for the system.
Therefore, we might expect some secondary source working against synchronized rotation, which might be indirect evidence
for dynamical perturbations of binary `A' on binary `B'. 

\section{High Resolution Imaging}
\label{sec:AO}

\subsection{Keck Adaptive Optics}
\label{sec:Keck}

To obtain better constraints on this quintuple system, we imaged both E1213 and E1234 on 2016 April 13 UT with the NIRC2 instrument (PI: Keith Matthews) on Keck II using natural guide star imaging with the narrow camera setting (10 milliarcseconds, `mas', pixel$^{-1}$). We used a three-point dither pattern to avoid NIRC2's noisier lower left quadrant and we calibrated the images and removed artifacts using dome flat fields and dark images.

\begin{figure}
\begin{center}
\includegraphics[width=0.48 \textwidth]{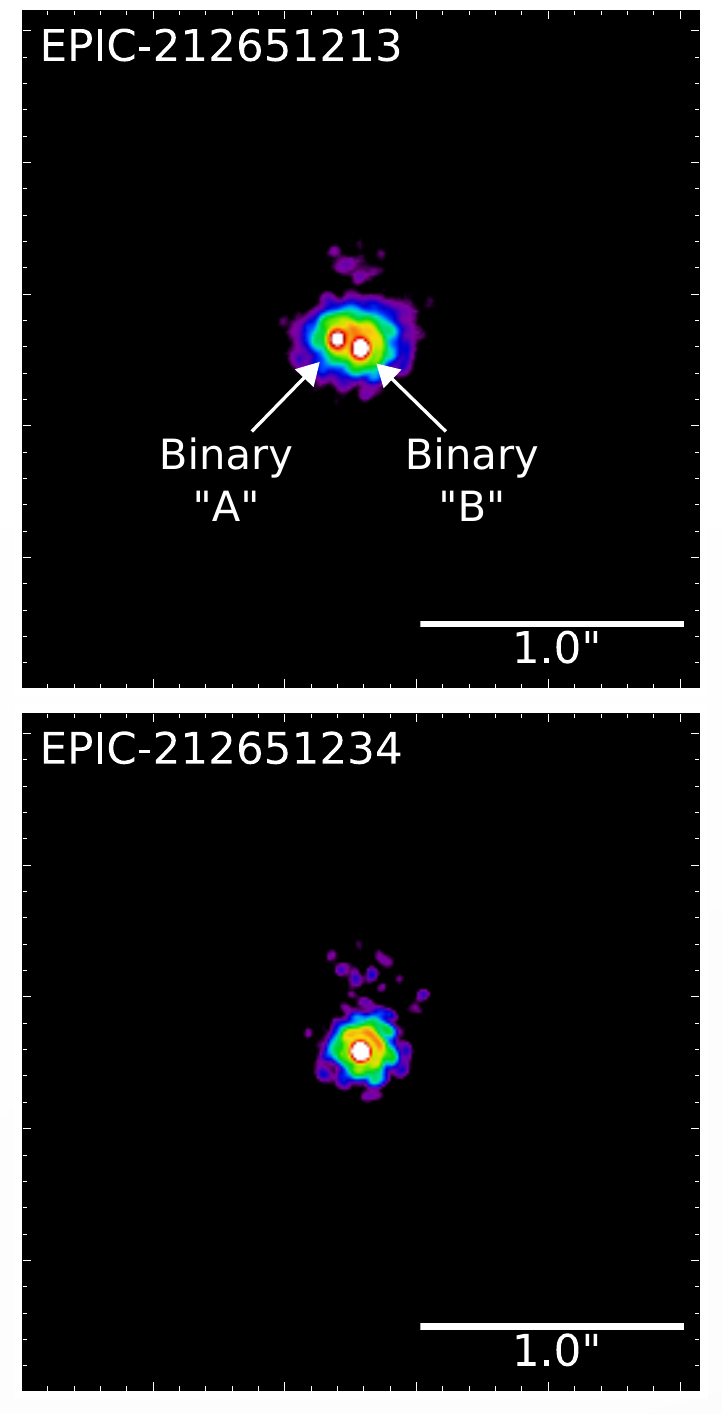}
\caption{Top panel: Keck-AO image in $K_s$-band of E1213.  This target clearly has two cores separated by $\sim$0.095$''$.  From this we conclude that the separation between binaries 'A' and 'B' is $\approx 25 \pm 5$ AU. Bottom panel: Star E1234 is unresolved at the $\sim$0.05$''$ level for stars of comparable magnitude. }
\label{fig:AO}
\end{center}
\end{figure}

We obtained 12 images of E1213 in the $K_s$ band (central wavelength 2.145 $\mu$m) for a total on-sky integration time of 300 seconds. Figure \ref{fig:AO} (top panel) shows a stacked image of this target, in which we see two peaks. For each calibrated frame, we fit a two-peak PSF to measure the flux ratio and on-sky separation. We model the PSF as a combined Moffat and Gaussian shape. The best-fit PSF was found over a circular area with a radius of 10 pixels around each star (the full width at half-maximum of the PSF was about 5 pixels). More details of the method can be found in Ngo et al. (2015). When computing the separation and position angle, we applied the new astrometric corrections from Service et al.~(2016) to account for the NIRC2 array's distortion and rotation. We find the $\Delta K_s$ for binary `A' $-$ binary `B' to be $0.64 \pm 0.03$, the separation to be $90.5 \pm 1.4$ mas, and the position angle of `A' relative to `B' to be $68.5 \pm 0.9$ deg E of N. We also took 6 images in J band (central wavelength 1.248 $\mu$m).  
We find the $\Delta J$ for binary `A' $-$ binary `B'  to be $0.48 \pm 0.01$.  The magnitude differences between the `A' and `B' binaries are summarized in Table \ref{tbl:ratios}.  

We also imaged E1234 with the same setup parameters and show the stacked image in the bottom panel of Figure \ref{fig:AO}. We do not see more than one light source, so we compute a 5-$\sigma$ contrast curve to put an upper limit on an unseen companion. To determine our limiting contrast over a range of distances, we divide the stacked image into a series of annuli centered on the star with a width of 5 pixels. Then, for every pixel in each annuli, we compute the sum of all neighboring pixels in a $5 \times 5$ box. The standard deviation of these values determines the $1\sigma$ contrast for that separation. The limiting magnitude is determined by dividing each limiting contrast by the flux in a $5 \times 5$ box centered on the primary star.  At separations of 0.05$''$, 0.10$''$, 0.15$''$, 0.20$''$, and 0.5$''$ from E1234, we rule out additional companions brighter than $\Delta K_s \simeq 0$, 1.4, 2.8, 3.9, and 6, magnitudes, respectively.

\begin{table}
\centering
\caption{Light Ratios of `A' to `B'}
\begin{tabular}{lc}
\hline
\hline
Waveband & $\Delta$ Magnitude (`A'-`B')\\
\hline
0.692 $\mu$m \ &  $0.55 \pm 0.1$$^a$  \\ 
0.880 $\mu$m  &  $0.45 \pm 0.1$$^a$  \\
J & $0.48 \pm 0.01$$^b$  \\
Ks &  $0.64 \pm 0.03$$^b$  \\ 
\hline
\end{tabular}
\label{tbl:ratios}

{\bf Notes.} (a) Based on the speckle images (see Sect.~\ref{sec:speckle}). (b) Based on the Keck AO imaging (see \ref{sec:Keck}).
\end{table}

\subsection{Speckle Interferometry}
\label{sec:speckle}
	
We performed speckle interferometric observations at the 3.5-m WIYN telescope located on Kitt Peak using the Differential Speckle Survey Instrument (see Horch et al.~2009; 2011).  This speckle camera provides simultaneous observations in two filters by employing a dichroic beam splitter and two identical electron-multiplying CCDs as the imagers. We observed E1213 and E1234 on 17 April 2016 UT obtaining 5 sets of 1000, 40-msec images for each star. The observations were performed simultaneously in narrow ``R'' and ``I'' bandpasses, where ``R''  and ``I'' have central wavelengths of 0.692 $\mu$m and 0.880 $\mu$m, respectively, and the corresponding FWHMs are 0.04 $\mu$m and 0.05 $\mu$m.  Our final reconstructed images and results of the observations were obtained using the full combined data set. The details of how we obtain, reduce, and analyze the speckle interferometric results are described in Howell et al.~(2011).

The speckle imaging also resolves the image of E1213 into two images that we associate with binaries `A' and `B' just as we found from the AO imaging (see Sect.~\ref{sec:speckle}).  The separation on the sky in the speckle image confirms the angular separation at $90.7 \,\,{\rm mas} \pm 1.5 \,\, {\rm mas}$ found in the Keck AO image.  The position angle of $68.8^\circ$ agrees with that determined with the Keck AO to within the speckle uncertainty of 1.3$^\circ$.  We use the speckle images to determine the ratio of the fluxes from binary `B' to binary `A', and find $1.66 \pm 0.16$ at 0.69 $\mu$m, and $1.51 \pm 0.15$ at 0.88 $\mu$m.  These ratios are summarized in Table \ref{tbl:ratios}.  When combined with the three mass functions that we measure for binaries `A' and `B', these flux ratios play an important role in determining the properties of the four stars (see Sect.~\ref{sec:MC_eval}).

\section{Photometric Distance Estimate}
\label{sec:dist}

We can estimate the distance to E1213 and E1234 from photometric parallax.  For E1213, comprised of binaries `A' and `B', we have good estimates of the masses, and evolutionary states of the four constituent stars, and hence their combined luminosities.  We estimate that stars 1 through 4 have a combined bolometric luminosity of $L_{\rm tot} \simeq 2.5 \pm 0.8 \, L_\odot$.  Since the three stars that dominate the luminosity budget are not far from solar-type stars, we take this ensemble of stars to be $1.0 \pm 0.3$ visual magnitudes brighter than the Sun (i.e., no significant bolometric correction), and thus it has $M_V \simeq 3.7 \pm 0.3$.  The visual magnitude of E1213 is $V \simeq 10.8$, and therefore the distance modulus to the source is 7.1 magnitudes.  Finally, this leads to a distance estimate of 260 pc.  When done within our Monte Carlo parameter estimation code (see Sect.~\ref{sec:MC_eval}), we find $d = 260 \pm 50$ pc.  

For the stellar image E1234, we estimate that this apparently single star has $T_{\rm eff} \simeq 4987 \pm 50$ K and $\log \, g \simeq  3.73 \pm 0.1$ (see Table \ref{tbl:decomp}).  Its location is superposed on evolution tracks in the $\log g - T_{\rm eff}$ plane in Fig.~\ref{fig:YY}, and one can readily see that it is substantially evolved.  From the location of E1234 in the HR diagram, and the use of the Yonsei-Yale evolution tracks (Yi et al.~2001), we estimate that its bolometric luminosity is $3.4 \pm 0.9 \, L_\odot$, for stars between 1.0 and 1.3 $M_\odot$. The apparent magnitude for this star is $V \simeq 11.1$, which roughly translates to $m_{\rm bol} \simeq 10.9$.  The distance modulus then comes out to be $7.5 \pm 0.3$, corresponding to a distance of $315 \pm 50$ pc. This distance estimate is therefore quite consistent with that found above from the brightness of binaries `A' and `B'.

\begin{table}
\centering
\caption{Inferences About the `C' Binary}
\begin{tabular}{lcc}
\hline
\hline
Parameter &
 Binary A & Binary B \\
\hline
$\gamma$ (km/s)$^a$ &  $-8.4 \pm 1.0$ & $-19.0 \pm 0.2$ \\ 
$K$ (km/s)$^b$ &  $\simeq 5.2 $  & $\simeq 5.5$ \\
$\left| \dot \gamma\right| $ (cm s$^{-2}$) & $< 0.01$ & $< 0.006$ \\
$\theta_{\rm AB}$ (arcsec)$^c$ &  $0.095$ & $0.095$ \\ 
$P_{\rm AB}$ (yr)$^d$ & $65 \pm 20$ & $65 \pm 20$ \\
$a_{\rm AB}$ (AU)$^d$ & $25 \pm 5$ & $25 \pm 5$ \\
\hline
\end{tabular}
\label{tbl:Cbinary}

{\bf Notes.} (a) Taken from Tables \ref{tbl:Abinary} and \ref{tbl:Bbinary}. (b) We apportioned the difference in $\gamma$ velocities inversely proportional to the masses of binary A (1.82 $M_\odot$) and binary B (1.73 $M_\odot$). (c) Based on the AO imaging. (d) Based on a Monte Carlo selection of binary inclination angles and phases (see Sect.~\ref{sec:binaryC}).
\end{table}

\section{Binary `C': Comprised of Binaries A \& B}
\label{sec:binaryC}

Having established the basic properties of binaries `A' and `B', we can now consider them as likely bound members of a higher-order binary (i.e., a quadruple star system) which we call `C'.  We know three quantitative things about the `C' binary: (1) The relative radial velocity of `A'  relative to `B' is $\simeq$ 10.6 km s$^{-1}$ as inferred from the difference in $\gamma_A$ and $\gamma_B$, hereafter `$\Delta \gamma_{\rm AB}$'.  (2) The projected angular separation between these two components is $\simeq 0.095''$, based on the imaging study (see Sect.~\ref{sec:AO}), which corresponds to a physical projected separation of $s \simeq 25 \pm 5$ AU at the estimated distance of $260 \pm 50$ pc.  (3) The relative radial acceleration of the two binaries with respect to each other is constrained to be $\lesssim 0.01$ cm s$^{-2}$ based on limits to $\dot \gamma$ for the `A' and `B' binaries (see Table \ref{tbl:Cbinary}).  We can use these three facts to set some interesting constraints on the orbital period and inclination of the `C' binary following the approach used in Lehmann et al.~(2016) for the case of the quadruple system KIC 7177553.  

Because we have only (i) the relative radial velocity between the two binaries, $\Delta \gamma_{\rm AB}$, (ii) a value for the projected physical separation, $s$, and (iii) an upper limit on the relative radial acceleration of the two binaries, we consider only the simple case of a circular orbit for the `C' binary as being illustrative.  These three quantities are related to the unknown orbital radius, $a_C$, inclination, $i_C$, and phase, $\phi_C$ by:
\begin{eqnarray}
s & \, = \, & a_C \sqrt{\cos^2{i_C} + \sin^2{\phi_C}\sin^2{i_C}} \\
\Delta\gamma_{\rm AB} & \, = \, & \sqrt{\frac{G M_C}{a_C}}\sin{\phi_C}\sin{i_C} \\
\dot \gamma_{\rm AB} & \, = \, & \frac{GM_C}{a_C^2} \, \cos \phi_C \sin i_C
\end{eqnarray}
where  $M_{\rm C}$ is the total system mass (i.e., the combined masses of binaries `A' and `B').  Note that $\phi_C$ is taken here to be zero at the time of superior conjunction, which is 90$^\circ$ from the definition used by Lehmann et al.~(2015).  Following Lehmann et al.~(2016), this unknown orbital phase can be eliminated from Eqns.~(8) and (9) to find a cubic expression for the orbital separation:
\begin{equation}
a_C^3 \left(\frac{\Delta\gamma^2}{G M_C}\right) + a_C^2 \cos^2 i_C- s^2 =0
\label{eqn:aC}
\end{equation}

In spite of the fact that we do not know the orbital inclination angle, $i_C$, we can still produce a probability 
distribution for $a_C$ (and hence $P_{\rm orb,C}$) via a Monte Carlo approach as follows.  For each realization of the system 
we choose a random inclination angle with respect to an isotropic set of orientations of the orbital angular 
momentum vector of the `C' binary.  In addition, because there is an uncertainty in the determination of $\Delta\gamma_{\rm AB}$ 
(see Tables\,\ref{tbl:Abinary} and \ref{tbl:Bbinary}), and a significant uncertainty in the projected orbital separation, $s$ 
(due largely to the uncertainty in distance to the target), we also choose specific 
realizations for those two quantities using a Gaussian random draw for both $\Delta\gamma_{\rm AB}$ and $s$.  
We then solve Eqn.~(\ref{eqn:aC}) for $a_C$, and we also store 
the corresponding value of $i_C$.  Finally, we restrict the inferred value of $a_C$ so as to yield a value for the radial acceleration (see Eqn.~10) that 
is lower than the observed upper limit cited in Table \ref{tbl:Cbinary}.  If any inclination angle leads to a non-physical solution of Eqn.~(\ref{eqn:aC}), 
or a radial acceleration that is too large, we discard that system.  

\begin{figure}
\begin{center}
\includegraphics[width=0.48 \textwidth]{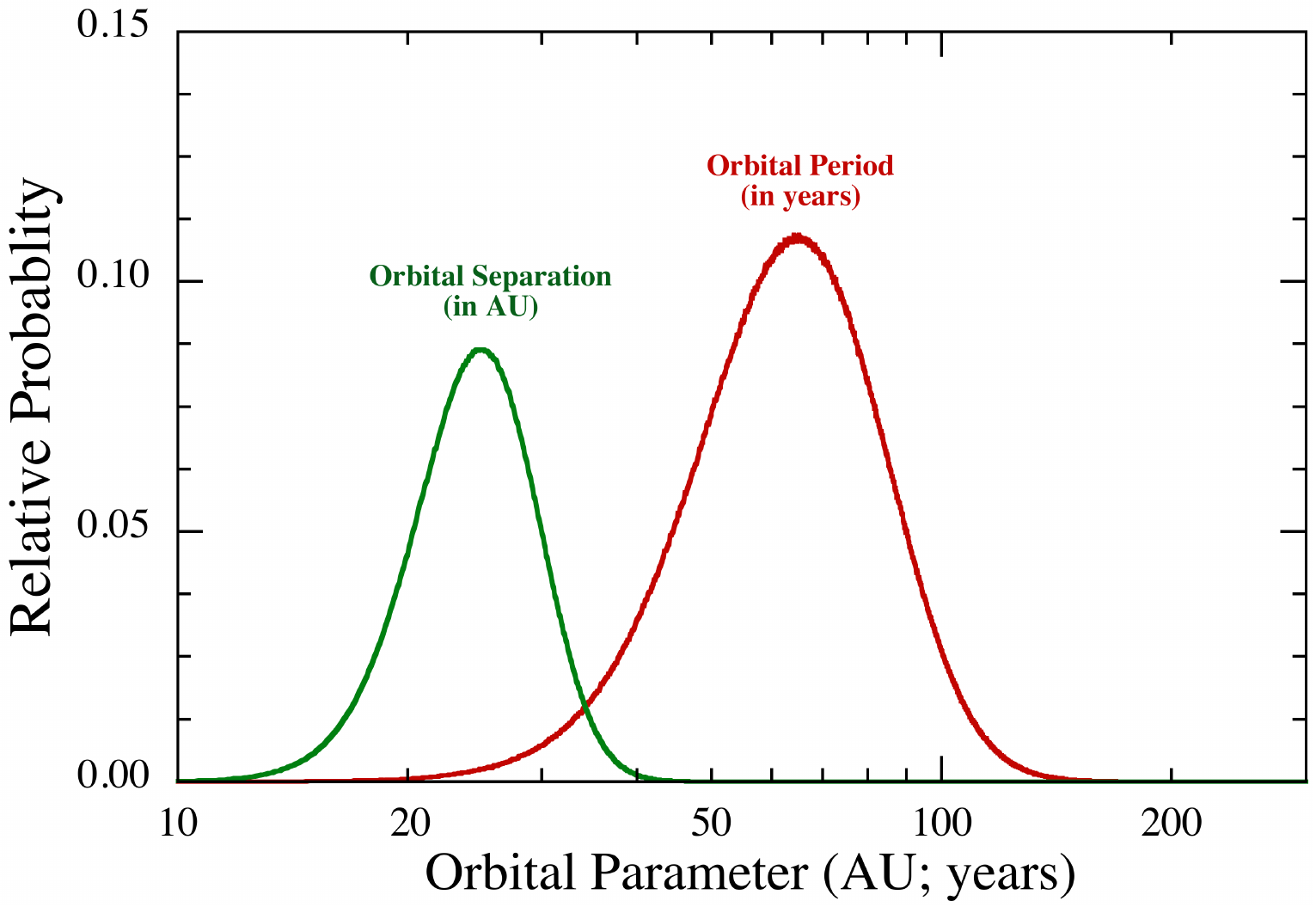}
\includegraphics[width=0.48 \textwidth]{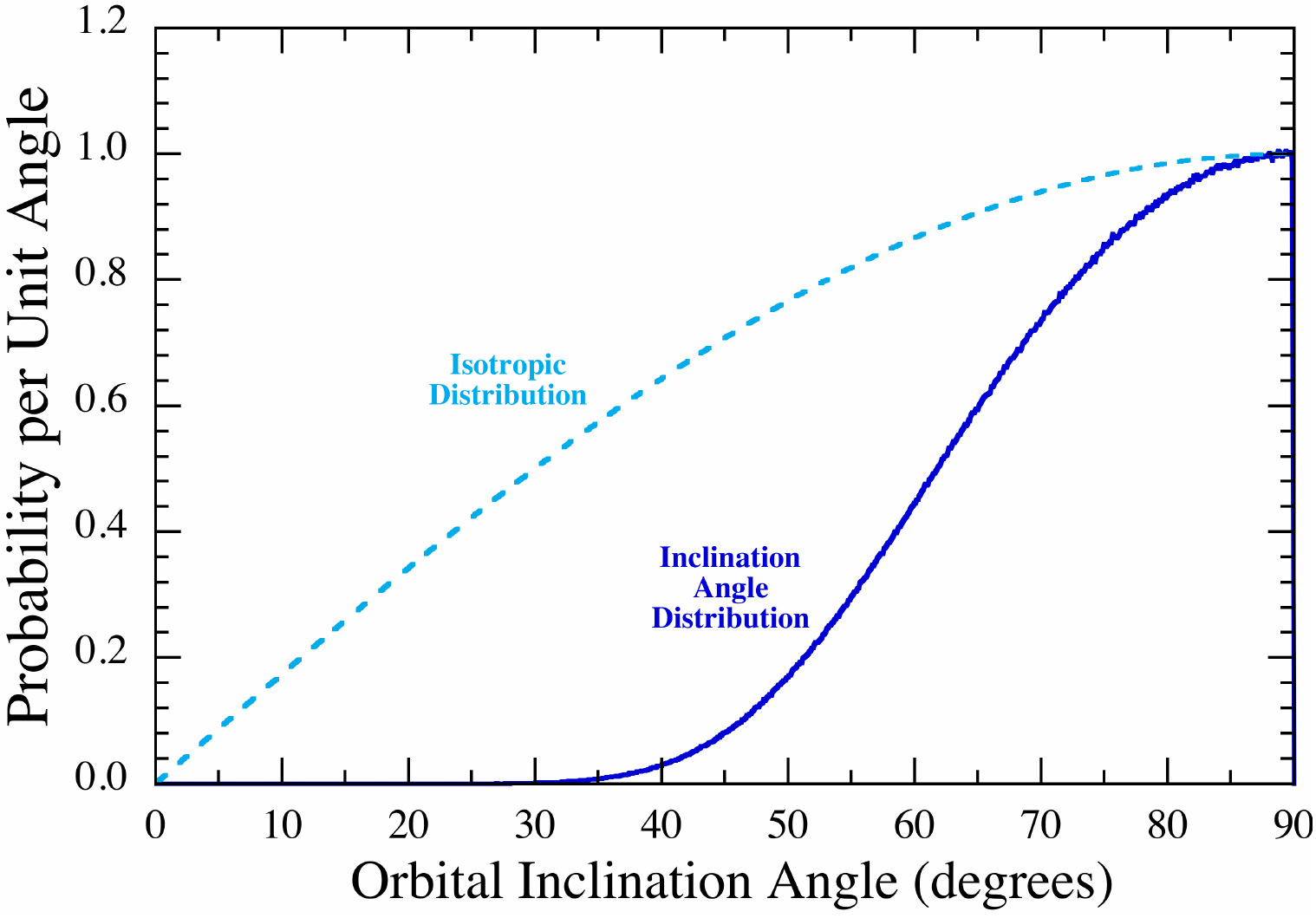}
\caption{{\em Top panel}: Orbital period distribution for binary `C' (orbital period in red; orbital separation in green).  {\em Bottom panel}:  Probability per unit orbital inclination angle for binary `C'.  The constraints are provided by $\Delta \gamma_{\rm AB}$, the upper limit on $\dot \gamma_{\rm AB}$, and the angular  separation of binaries `A' and `B' (see summary in Table \ref{tbl:Cbinary}). }
\label{fig:per_dist}
\end{center}
\end{figure}

We repeat this process $10^8$ times to produce our distributions of $a_C$, $P_{\rm orb, C}$ and $i_C$.  Probability distributions for $P_{\rm orb,C}$, $a_C$, and $i_C$ are shown in Fig.~\ref{fig:per_dist}.  The orbital period of binary `C' is $65 \pm 20$ years (1-$\sigma$ uncertainties).  The corresponding orbital separation is $25 \pm 5$ AU.  According to the orbital inclination angle distribution, values of $i_C \gtrsim 50^\circ$ are strongly favored (see bottom panel of Fig.~\ref{fig:per_dist}).  Such orbital periods are sufficiently short that the prospects for near-term detections of the accelerated
motions of binaries `A' and `B' are highly promising.

\begin{figure}
\begin{center}
\includegraphics[width=0.48 \textwidth]{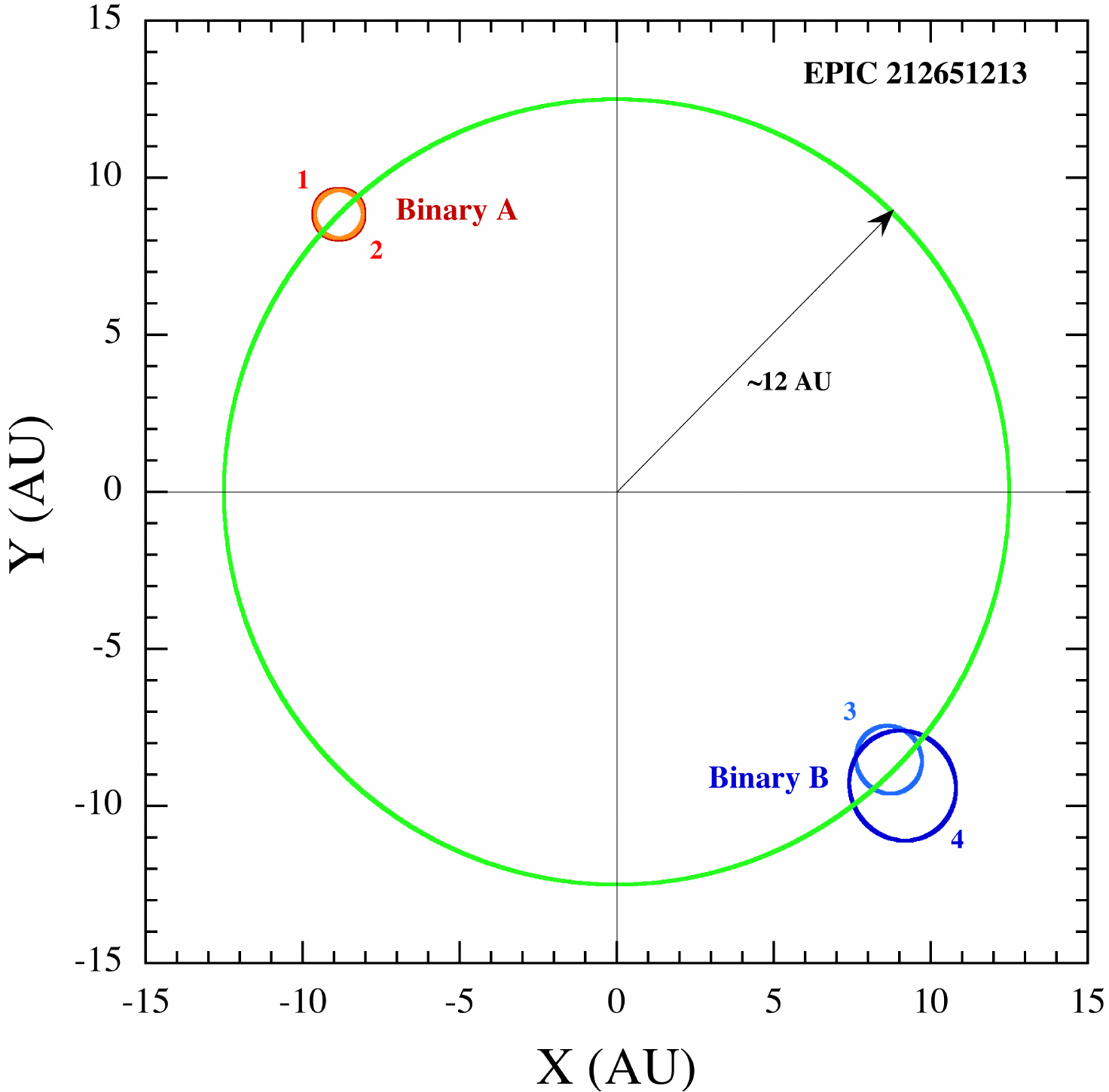}
\caption{Schematic of the quadruple comprised of binaries `A' and `B' viewed from the orbital pole.  The outer orbit of binary `A' around binary `B' is drawn to scale, the orbit is assumed circular, and the masses of the two binaries are taken to be equal (to within their uncertainties).  The orbits of stars 1 and 2 in the circular binary `A' and stars 3 and 4 in eccentric binary `B' are drawn to the correct shape and size relative to each other; however, their absolute size has been artificially scaled up by a factor of $\sim$20 so that the orbits can be resolved.  The relative orbital speed of the `A' and `B' binaries would be about 10 km s$^{-1}$ and the orbital period would be $\sim$65 years.}
\label{fig:scheme}
\end{center}
\end{figure}

Given that the projected separation of binaries `A' and `B' on the sky is very similar (within the uncertainties) to the peak value of $a_C$ in the distribution of orbital separation (see Fig.~\ref{fig:per_dist}), i.e., $s \simeq a_C$ in Eqn.~(8), this implies that $\phi_C \simeq 90^\circ$ or 270$^\circ$.

Repeating this exercise while allowing for eccentric orbits tends to broaden the distributions by a factor of about 2.  We examined this question extensively in Lehmann et al.~(2016).  This would not change the conclusions reached here in a material way except to increase the uncertainties.

A sketch of the `C' binary is shown in Fig.~\ref{fig:scheme}.  The outer orbit of binary `A' around binary `B' is drawn to scale, and the orbit is assumed circular.  The orbits of stars 1 and 2 in the circular binary `A' and stars 3 and 4 in eccentric binary `B' are drawn to the correct shape and size relative to each other; however, their absolute size has been artificially scaled up by a factor of $\sim$20 so that the orbits can be resolved.

Since we are likely viewing binary `C' with large inclination angles and $\phi_C \sim 90^\circ$, we can reduce Eqns.~(8) and (9) to:
\begin{eqnarray}
\frac{\Delta s[\delta \phi_C]}{d_{\rm pc}}  \sim 45 \, \delta \phi_C^2 ~{\rm mas} \\
\Delta \gamma_{\rm AB}[\delta \phi_C] \sim 5 \, \delta \phi_C^2 ~{\rm km~s}^{-1} 
\end{eqnarray}
where $d$ is the distance to the target, $\Delta s$ is the change in projected physical separation (in AU), and $\delta \phi_C$ is the advance in the orbital phase (in radians) over the next few years.  Future RV and high-resolution imaging observations of this object should readily achieve accuracies of 1.5 mas in terms of the angular separation of binaries `A' and `B', and 1 km s$^{-1}$ for the RV measurements. Solving for the orbital phase changes required to make a detection of orbital motion in binary `C' we find $\delta \phi_C \simeq 0.03$ and 0.07 orbital cycles for the high-resolution imaging and the RV observations, respectively.  These orbital phase advances correspond to $\sim$2 and 4.5 years, respectively.  Thus, we advocate further RV and high-resolution imaging of this object starting a couple of years from now.

\section{E1213 and E1234 as a Binary Pair}
\label{sec:Dbinary}

The three facts we have to work with concerning the relationship between objects E1213 (comprised of a quadruple star system) and E1234 (which appears single both in imaging and spectroscopy), are the following.  First, the projected separation of the two stars is $10.56''$ on the sky. For a distance of about 260 pc, this corresponds to a projected physical separation of about 2800 AU or 0.013 pc.  This is well within the allowed separation for stars to remain bound pairs without being tidally disrupted in their passage around the Galaxy (Jiang \& Tremaine 2010).  Second, the radial velocity of the center of mass of the `C' binary (actually a quadruple system) is $-13.6$ km s$^{-1}$ while the radial velocity of E1234 is -15.0 km s$^{-1}$, both with an uncertainty of $\sim$1 km s$^{-1}$.  These are suggestively close to having the same value.  Third, the magnitude of the difference in the two proper motion vectors is only $3.4 \pm 2.3$ mas yr$^{-1}$.  

Combining these three facts leads to the quite plausible, even highly likely, notion that E1213 and E1234 are themselves a gravitationally bound binary pair.  We refer to this as the `D' binary.  The angular, velocity, and time scales are too long to hope to see any changes in this binary over the next decades.  

\section{E1234 as a Single Star}
\label{sec:E1234}

The rms scatter of the RVs for E1234 within the combined TRES and KO sets is $\sim$100 m s$^{-1}$, while for the TLS observations by themselves, the rms scatter is similar (see Table \ref{tbl:RVD}).   However, there is a systematic offset between these two sets of $\sim$900 m s$^{-1}$.  If we subtract off that constant offset from the three TLS RV measurements of E1234, the overall rms spread in the RV measurements remains at 100 m s$^{-1}$.  We are unable to account for the systematic shift between observatories, which has minimal effect on the evaluation of the orbital parameters in binaries `A' and `B'.  However, for estimating an upper limit to the projected radial acceleration of a possible binary in E1234 this discrepancy matters, and we adopt an rms radial velocity of 100 m s$^{-1}$ for this purpose.  In turn, this yields an upper limit to the projected radial acceleration of $\sim$0.003 cm sec$^{-2}$.  

In addition, the angular separation of any internal stellar pairs is $\lesssim 0.05''$ based on the AO and speckle imaging (see Sect.~\ref{sec:AO}).  

The limits on angular separation and projected radial accelerations suggest that any binary stellar pair within E1234 has a physical separation, $a$ constrained by:
\begin{equation}
20 \sqrt{\sin i \, \cos \phi}  \, \lesssim ~a~ \lesssim \, 13 \left(\cos^2 i + \sin^2 \phi \, \sin^2 i \right)^{-1/2} {\rm AU} 
\label{eqn:E1234}
\end{equation}
where $i$ and $\phi$ are the orbital inclination angle and phase, respectively, of any hypothetical binary in E1234 (as in Eqns.~8 and 10). Here we have assumed a total binary mass of $2 \, M_\odot$.

Interestingly, unless the orbital phase or inclination angle are somewhat finely tuned, the above constraints already suggest that any putative binary in E1234 must have an orbital separation not too far from $\sim$16 AU.  If it is much wider it would have been resolved with AO or speckle imaging, and if much closer the acceleration (i.e., change in $\gamma$ velocity) would likely have been detected over the 72-day interval of the RV measurements.

To push this argument a bit further, we took advantage of TRES's 15 m s$^{-1}$ instrumental stability.  Relative RVs were derived by cross-correlating each observed spectrum, order-by-order, against the strongest observed spectrum over the wavelength range 0.435-0.628 $\mu$m.  These yield significantly tighter constraints on {\em changes} in velocity over the course of the observations than the absolute velocities given in Table \ref{tbl:RVD}.  With these multi-order velocities, we can set a limit of $\approx$ 30 m s$^{-1}$ on the constancy of the radial velocity for E1234.  In turn, this lower limit increases the coefficient on the left side of Eqn.~(\ref{eqn:E1234}) to a value of $\sim$35. This constraint effectively eliminates nearly all viable solutions for the semi-major axis of any putative circular binary.  All this is in keeping with our assertion that image E1234 is that of a single star.

Finally, we note that if the location of star E1234 in the $\log \,g - T_{\rm eff}$ is as shown in Fig.~\ref{fig:YY}, it is significantly evolved, and would have a mass of $\lesssim 1.3 \,M_\odot$ and an age $\gtrsim 4.4$ Gyr.  If we take E1213 and E1234 to be coeval, this sets the same lower limit to the age of the two binaries in E1213.

\section{Summary and Conclusions}
\label{sec:concl}

We have presented an unusual quintuple star system consisting of 5 stars arranged as a hierarchical quadruple orbited by a single star.  The K2 data show that both binaries comprising the quadruple are eclipsing (Fig.~\ref{fig:folds}).  We have measured the spectroscopic orbits of 3 of the 4 stars in the two binaries (Fig.~\ref{fig:RVs}).  The 5-day binary is highly circular, while the 13-day binary is eccentric with $e_B \simeq 0.32$ (Fig.~\ref{fig:RVs} and Tables \ref{tbl:Abinary} and \ref{tbl:Bbinary}).  The separation of the two binaries is resolved by both Keck AO imaging and WIYN speckle interferometry at 90 mas (Fig.~\ref{fig:AO}).  The flux ratio of the two binaries was thereby measured quite accurately at four different wavelengths with the AO and speckle imaging (Table \ref{tbl:ratios}).  By combining the three measured mass functions with the flux ratio of binary `B' to `A', and utilizing stellar evolution models, we have managed to extract reasonably accurate stellar parameters for all four stars in the two binaries (Fig.~\ref{fig:dist}).  

We have shown that if the RV and/or high-resolution imaging of this object are repeated 2-3 years from now, there is a good chance that changes in the motion of binary `A' with respect to binary `B' (i.e., within the binary `C' system) will be detected.  Additionally, in regard to detecting motion within the `C' binary, we note that the eclipses of binary `A' are deep enough to easily follow from ground-based observations.  If the eclipses can be tracked with an accuracy of minutes, then we should be able to detect significant eclipse timing variations within a few years.  This is physically equivalent to measuring $\dot \gamma_{\rm AB}$ via RV observations.

After analyzing and evaluating the quadruple system E1213, we argued that the stellar image E1234 is itself a single star (see Sect.~\ref{sec:E1234}), and is very likely bound to the quadruple system (see Sect.~\ref{sec:Dbinary}). 

This system is one of only a relative handful of known quadruple or higher multiplicity stellar systems where the radial velocities have been measured for both binaries, and where, additionally, both binaries are eclipsing.  It has a short enough outer orbital period so that motion of that longer period system can be detected within a few years.

\acknowledgements 

\vspace{0.3cm}

We are grateful to Mark Everett for help with the WIYN observations.  We thank Alan Levine for helpful discussions about this system.  Some of the data presented in this paper were obtained from the Mikulski Archive for Space Telescopes (MAST). STScI is operated by the Association of Universities for Research in Astronomy, Inc., under NASA contract NAS5-26555. Support for MAST for non-HST data is provided by the NASA Office of Space Science via grant NNX09AF08G and by other grants and contracts.  Based, in part, on data from CMC15 Data Access Service at CAB (INTA-CSIC).  This work was based on observations at the W. M. Keck Observatory granted by the California Institute of Technology. We thank the observers who contributed to the measurements reported here and acknowledge the efforts of the Keck Observatory staff. We extend special thanks to those of Hawaiian ancestry on whose sacred mountain of Mauna Kea we are privileged to be guests. A.\,V.~is supported by the NSF Graduate Research Fellowship, Grant No.~DGE 1144152.  E.\,H. is grateful for support from NASA's Ames Research Center that allowed him to participate in the speckle observations and analysis.  B.\,K. gratefully acknowledges the support provided by the Turkish Scientific and Technical Research Council  (T\"UB\.ITAK-112T766 and T\"UB\.ITAK-B\.IDEP 2219). K.\,P. was supported by the Croatian HRZZ grant 2014-09-8656.  \'A.\,S. acknowledges the financial support of the Hungarian NKFIH Grant K-115709 and the J\'anos Bolyai Research Scholarship of the Hungarian Academy of Sciences. T.\,B. and \'A.\,S. acknowledge the financial support of the NKFIH Grant OTKA K-113117.  The Konkoly observations were supported by the Lend\"ulet grant LP2012-31 of the Hungarian Academy of Sciences.


\begin{thebibliography}

\bibitem[Ahn et al.(2012)]{Ahn} Ahn, C.P., Alexandroff, R., Prieto, C.A., et al. 2012, ApJS, 203, 21

\bibitem[Asplund(2009)]{Asplund} Asplund, M., Grevesse, N., Sauval, A.J., \& Scott, P. 2009, ARA\&A, 47, 481

\bibitem[Barstow et al.(2001)]{Barstow} Barstow, M.A., Bond, H.E., Burleigh, M.R., \& Holberg, J.B. 2001, MNRAS, 322, 891

\bibitem[Batalha et al.(2011)]{2011ApJ...729...27B} Batalha, N.~M., Borucki, W.~J., Bryson, S.~T., et al.\ 2011, \apj, 729, 27 

\bibitem[Borkovits et al.(2015)]{Borko15} Borkovits, T., Rappaport, S., Hajdu, T., \& Sztakovics, J. 2015, MNRAS, 448, 946

\bibitem[Borkovits et al.(2016)]{Borko16} Borkovits, T., Hajdu, T., Sztakovics, J., Rappaport, S., Levine, A., B\'ir\'o, I.B., \& Klagyivik, P. 2016, MNRAS, 455, 4136

\bibitem[Borucki et al.(2010)]{Borucki} Borucki, W.J., Koch, D., Basri, G., et al. 2010, Sci, 327, 977

\bibitem[Buchhave et al.(2010)]{Buchhave} Buchhave, L.A., Bakos, G.\'A., Hartman, J.D., et al. 2010, ApJ, 720, 1118

\bibitem[Buchhave et al.(2012)]{Buchhave12} Buchhave, L. A., Latham, D. W., Johansen, A., et al. 2012, Nature, 486, 375

\bibitem[Conroy et al.(2014)]{Conroy} Conroy, K.E., Pr\v{s}a, A., Stassun, K.G., Orosz, J.A., Fabrycky, D.C., \& Welsh, W.F. 2014, AJ, 147, 45

\bibitem[Cutri et al.(2013)]{Cutri} Cutri, R.M., Wright, E.L., Conrow, T., et al.~2013, wise.rept, 1C.

\bibitem[Derekas et al.(2016)]{Derekas} Derekas, A., Plachy, E., Moln\'ar, L., S\'odor, \'A., Benk\H o, J. M., Szabados, L., Bogn\'ar, Zs., Cs\'ak, B., Szab\'o, Gy. M., Szab\'o, R., P\'al, A.  2016, submitted to MNRAS

\bibitem[Di Folco et al.(2014)]{DiFolco} Di Folco, E., Dutrey, A., Le Bouquin, J.-B., et al. 2014, A\&A, 565, 2

\bibitem[Hadrava(1995)]{Hadrava} Hadrava, P. 1995, A\&AS, 114, 393

\bibitem[Hadrava(2004)]{Hadrava04} Hadrava, P. 2004, in ``Spectroscopically and Spatially Resolving the Components of the Close Binary Stars", ASPC, eds. R.W. Hilditch, H. Hensberge, \& K. Pavlovski, 318, 86

\bibitem[Horch et al.(2009)]{Horch09} Horch, E.P., Veillette, D.R., Baena Gall\'{e}, R., Shah, S.C., O'Rielly, G.V., \& van Altena, W. 2009, AJ, 137, 5057

\bibitem[Horch et al.(2011)]{Horch} Horch, E. P., Gomez, S. C., \& Sherry, W. H. et al. 2011, AJ, 141, 45

\bibitem[Howell et al.(2011)]{Howell11} Howell, S. B., Everett, M. E., Sherry, W., Horch, E., \& Ciardi, D. R. 2011, AJ, 142, 19

\bibitem[Howell et al.(2014)]{Howell14} Howell, S.B., Sobeck, C., Haas, M., et al. 2014, PASP, 126, 398

\bibitem[Huber et al.(2015)]{Huber} Huber, D., Bryson, S.T., Haas, M.R., et al. 2015, arXiv:1512.02643.

\bibitem[Hut (1981)]{hut81} Hut, P., 1981, \aap, 99, 126

\bibitem[Ilijic (2004)]{Ilijic04} Ilijic, S. 2004, ``Spectroscopically and Spatially Resolving the Components of the Close Binary Stars'', ASPC, eds. R.W. Hilditch, H. Hensberge, \& K. Pavlovski, 318, 107

\bibitem[Jiang \& Tremaine (2010)]{Jiang} Jiang, Y.-F. \& Tremaine, S. 2010, MNRAS, 401, 977

\bibitem[Kirk et al.(2016)]{Kirk} Kirk, B., Conroy, K., Prs\v{a}, A., et al. 2016, AJ, 151, 68.

\bibitem[Kupka et al.(1999)]{Kupka} Kupka, F, Piskunov, N., Ryabchikova, T.A., Stemples, H.C., \& Weiss, W.W. 1999, in ``VALD-2: Progress of the Vienna Atomic Line Data Base'', 138, 119

\bibitem[Kurucz(1992)]{Kurucz:1992} Kurucz, R.~L.\ 1992, The Stellar Populations of Galaxies: Proceedings of the 149th Symposium of the International Astronomical Union, held in Angra dos Reis, Brazil, August 5-9, 1991. Eds.~B. Barbuy \& A.~Renzini, International Astronomical Union. Symposium no. 149, (Kluwer Academic Publishers; Dordrecht), 1992., p.~225

\bibitem[LaCourse et al.(2015)]{LaCourse} LaCourse, D.M., Jek, K.J., Jacobs, T., et al. 2015, MNRAS, 452, 3561

\bibitem[Lehmann et al.(2011)]{Lehmann} Lehmann, H., Tkachenko, A., Semaan, T., Guti{\'e}rrez-Soto, J., Smalley, B., Briquet, M., Shulyak, D., Tsymbal, V., \& De Cat, P. 2011, A\&A, 526, 124

\bibitem[Lehmann et al.(20012)]{Lehmann12} Lehmann, H., Zechmeister, M., Dreizler, S., Schuh, S., \& Kanzler, R. 2012, A\&A, 541, 105L

\bibitem[Lehmann et al.(2016)]{Lehmann16} Lehmann, H., Borkovits, T., Rappaport, S., Ngo, H, Mawet, D., Csizmadia, Sz., Forg\'acs-Dajka, E. 2016, ApJ, 819, 33.

\bibitem[Lohr et al.(2015)]{Lohr} Lohr, M.E., Norton, A.J., Gillen, E., Busuttil, R., Kolb, U.C., Aigrain, S., McQuillan, A., Hodgkin, S.T., \& Gonz\'alez, E. 2015, A\&A, 578, 103L

\bibitem[Matijevic et al.(2012)]{Matijevic} Matijevi\v{c}, G., Pr\v{s}a, A., Orosz, J.A., Welsh, W.F., Bloemen, S., \& Barclay, T. 2012, AJ, 143, 123

\bibitem[\protect\citeauthoryear{Moreno, Koenigsberger, Harrington}{2008}]{morenoetal11}
Moreno, E., Koenigsberger, G., Harrington, D. M., 2011, \aap, 528, A48

\bibitem[Ngo et al.(2015)]{Ngo} Ngo, H., Knutson, H. A., Hinkley, S., Crepp, J. R., Bechter, E. B., Batygin, K., Howard, A. W., Johnson, J. A., Morton, T. D., Muirhead, P. S. 2015, ApJ, 800, 138

\bibitem[Pr\v{s}a \& Zwitter(2005)]{Phoebe} Pr\v{s}a \& Zwitter 2005, ApJ, 628, 426

\bibitem[Pr\v{s}a et al.(2016)]{quint} Pr\v{s}a, A., et al., 2016 in preparation

\bibitem[Raghavan et al.(2009)]{Raghavan} Raghavan, D., McAlister, H.A., Torres, G., et al. 2009, ApJ, 690, 394

\bibitem[Rappaport et al.(2012)]{Rappaport12} Rappaport, S., Deck, K., Levine, A., Borkovits, T., Carter, J., El Mellah, I., Sanchis-Ojeda, R., \& Kalomeni, B. 2013, ApJ, 768, 33.

\bibitem[Schutz et al.(2011)]{Schutz} Sch\"utz, O., Meeus, G., Carmona, A., Juh\'asz, A., \& Sterik, M.F.  2011, A\&A, 533, 54

\bibitem[Service et al.(2016)]{Service} Service, M., Lu, J.R. , Campbell, R., Sitarski, B.N., Ghez, A.M., Anderson, J. 2016, submitted to PASP

\bibitem[Shibahashi \& Kurtz(2012)]{TheThing} Shibahashi, H., \& Kurtz, D.W. 2012, MNRAS, 422, 738

\bibitem[Simon \& Sturm(1994)]{Simon} Simon, K.P., \& Sturm, E. 1994, A\&A, 281, 286

\bibitem[Skrutskie et al.(2006)]{Skrutskie} Skrutskie, M.F., Cutri, R.M., Stiening, R., et al. 2006, AJ, 131, 1163.

\bibitem[Slawson et al.(2011)]{Slawson} Slawson, R.W., Pr\v{s}a, A., Welsh, W.F. 2011, AJ, 142, 160

\bibitem[Shulyak et al.(2004)]{Shulyak} Shulyak, D., Tsymbal, V., Ryabchikova, T., St{\"u}tz, C. \& Weiss, W. W. 2004, A\&A, 428, 993

\bibitem[Still \& Barclay(2012)]{Still} Still, M., \& Barclay, T. 2012, ascl.soft.08004

\bibitem[Szentgyorgyi \& Furesz (2007)]{Szentgyorgyi} Szentgyorgyi, A.-H., \& Fur\'esz, G. 2007, RMxAC, 28, 1298

\bibitem[Tkachenko et al.(2012)]{Tkachenko} Tkachenko, A., Lehmann, H., Smalley, B., Debosscher, J., \& Aerts, C. 2012, MNRAS, 422, 2960

\bibitem[Tokovinin (2016)]{Tokovinin} Tokovinin, A. 2016, arXiv, 160406399

\bibitem[Torres (2006)]{Torres} Torres, G. 2006, AJ, 131,1702

\bibitem[Tsymbal(1996)]{Tsymbal} Tsymbal, V. 1996, in ``M.A.S.S., Model Atmospheres and Spectrum Synthesis'', ASPC, 108, 198

\bibitem[Yi et al.(2001)]{Yi}  Yi, S., Demarque, P., Kim, Y.-C., et al. 2001, ApJS, 136, 417

\bibitem[Zacharias et al.(2013)]{UCAC4} Zacharias, N., Finch, C.T., Girard, T.M., Henden, A., Bartlett, J.L., Monet, D.G., \& Zacharias, M.I. 2013, ApJS, 145, 44

\bibitem[\protect\citeauthoryear{Zahn}{2008}]{zahn08}
Zahn, J.-P., 2008, in Tidal Effects in Stars, Planets and Disks, ed. M.-J. Goupil, \& J.-P. Zahn, EAS Pub. Ser., 29, 67

\bibitem[Zasche \& Uhlar(2013)]{V994Her} Zasche, P., \& Uhla\v{r}, R. 2016, A\&A, 588, 121


\end{thebibliography}
\end{document}